\documentclass[times,astrobib,amssymb,usenatbib]{mn2e}
\usepackage{graphicx}
\usepackage{txfonts}
\usepackage{natbib}
\usepackage{supertabular}
\usepackage{longtable}
\usepackage{url}
\usepackage{dcolumn}
\usepackage{placeins}
\usepackage{multirow}
\begin{document}
\newcommand{\sqcm}{cm$^{-2}$}  
\newcommand{\lya}{Ly-$\alpha$}
\newcommand{\lyb}{Ly-$\beta$}
\newcommand{\lyg}{Ly-$\gamma$}
\newcommand{\heo}{\mbox{He\,{\sc i}}}
\newcommand{\he}{\mbox{He\,{\sc ii}}}
\newcommand{\hi}{\mbox{H\,{\sc i}}}
\newcommand{\hw}{\mbox{H\,{\sc ii}}}
\newcommand{\os}{\mbox{O\,{\sc vi}}}
\newcommand{\cf}{\mbox{C\,{\sc iv}}}
\newcommand{\cto}{\mbox{C\,{\sc ii}}}
\newcommand{\ct}{\mbox{C\,{\sc iii}}}
\newcommand{\sit}{\mbox{Si\,{\sc iii}}}
\newcommand{\sif}{\mbox{Si\,{\sc iv}}}
\newcommand{\sito}{\mbox{Si\,{\sc ii1260}}}
\newcommand{\sia}{\mbox{Si\,{\sc ii1526}}}
\newcommand{\nf}{\mbox{N\,{\sc v}}}
\newcommand{\nt}{\mbox{N\,{\sc iii}}}
\newcommand{\zabs}{$z_{\rm abs}$}
\newcommand{\zqso}{$z_{\rm qso}$}
\newcommand{\subHe}{_{\it HeII}}
\newcommand{\subH}{_{\it HI}}
\newcommand{\subHLy}{_{\it H Ly}}
\newcommand{\degree}{\ensuremath{^\circ}}
%...............................................................................
\newcommand{\lapp}{\mbox{\raisebox{-0.3em}{$\stackrel{\textstyle <}{\sim}$}}}
\newcommand{\gapp}{\mbox{\raisebox{-0.3em}{$\stackrel{\textstyle >}{\sim}$}}}
\newcommand{\be}{\begin{equation}}
\newcommand{\en}{\end{equation}}
\newcommand{\di}{\displaystyle}
\def\tworule{\noalign{\medskip\hrule\smallskip\hrule\medskip}} %double rule.%
\def\onerule{\noalign{\medskip\hrule\medskip}} %single rule.%
\def\bl{\par\vskip 12pt\noindent}
\def\bll{\par\vskip 24pt\noindent}
\def\blll{\par\vskip 36pt\noindent}
\def\rot{\mathop{\rm rot}\nolimits}
\def\alf{$\alpha$}
\def\refff{\leftskip20pt\parindent-20pt\parskip4pt}
\def\zabs{$z_{\rm abs}$}
\def\zem{$z_{\rm em}$~}
\def\mgii{Mg\,{\sc ii}~}
\def\feiia{Fe\,{\sc ii}$\lambda$2600}
\def\mgia{Mg\,{\sc i}$\lambda$2852}
\def\mgiia{Mg\,{\sc ii}$\lambda$2796}
\def\mgiib{Mg\,{\sc ii}$\lambda$2803}
\def\mgiiab{Mg\,{\sc ii}$\lambda\lambda$2796,2803}
\def\wobs{$w_{\rm obs}$}
\def\kms{km~s$^{-1}$}
\def\chisq{$\chi^{2}$}
%========================================================================================

\title[He~{\sc ii} to H~{\sc i} ratio in the IGM]{Revisiting the He~{\sc ii} to H~{\sc i} ratio in 
       the Intergalactic Medium }
\author[S. Muzahid, R. Srianand, P. Petitjean]{S. Muzahid$^{1}$\thanks{E-mail: sowgat@iucaa.ernet.in},
R. Srianand$^{1}$, P. Petitjean$^{2}$ \\
$^{1}$ Inter-University Centre for Astronomy and Astrophysics, Post Bag 4, 
Ganeshkhind, Pune 411\,007, India \\
$^{2}$ Universit\'e Paris 6, UMR 7095, Institut d'Astrophysique de Paris-CNRS, 
98bis Boulevard Arago, 75014 Paris, France \\
}
\date{Accepted. Received; in original form }

%\pagerange{\pageref{firstpage}--\pageref{lastpage}} \pubyear{2010}

\maketitle
\label{firstpage}
%========================= ABSTRACT ==============================================================

\begin {abstract}
We estimate the He~{\sc ii} to H~{\sc i} column density ratio, $\eta = N(\he)/N(\hi)$,
in the intergalactic medium towards the high redshift ($z_{\rm em}$~=~2.885) 
bright quasar QSO HE~2347$-$4342 using Voigt-profile fitting of the H~{\sc i} 
transitions in the Lyman series and the He~{\sc ii} Lyman-$\alpha$ transition 
as observed by the {\sl FUSE} satellite. In agreement with previous studies, 
we find that $\eta > 50$ in most of the Lyman-$\alpha$ forest except in four 
regions where it is much smaller ($\eta \sim 10-20$) and therefore inconsistent 
with photo-ionization by the UV background flux. 
We detect O~{\sc vi} and C~{\sc iv}  absorption lines associated with two 
of these regions ($z_{\rm abs}$~=~2.6346 and 2.6498). 
We show that if we constrain the fit of the H~{\sc i} and/or He~{\sc ii} absorption 
profiles with the presence of metal components, we can accommodate $\eta$ values in 
the range 15-100 in these systems assuming broadening is intermediate between pure 
thermal and pure turbulent.
While simple photo-ionization models reproduce the  observed  
$N$(O~{\sc vi})/$N$(C~{\sc iv}) ratio, they fail
to produce low $\eta$ values contrary to models with high temperature (i.e T $\ge 10^5$~K).  
The Doppler parameters measured for different species suggest a multiphase nature of the
absorbing regions. Therefore, if low $\eta$ values were to be confirmed, we would favor a 
multi-phase model in which most of the gas is at high temperature ($>$ 10$^5$~K) 
but the metals and in particular C~{\sc iv}  are due to lower temperature ($\sim$ 
few $10^4$ K) photo-ionized gas.  
\end {abstract}

%========================== KEY WORDS ========================================================

\begin{keywords}
galaxies: quasar: absorption line -- quasar: individual(HE~2347$-$4342) -- galaxies: intergalactic 
medium
\end{keywords}

%=========================== INTRODUCTION ===================================================

\section{Introduction}

The presence of metals in the H~{\sc i} Lyman-$\alpha$ forest at optical depths 
$\tau_{\rm Ly\alpha}\ge 1$, detected through C~{\sc iv} and O~{\sc vi} absorption 
lines seen in QSO spectra, is now well established
\citep[see][]{Songaila96,Bergeron02,Simcoe04}.
Observations are consistent with an average carbon metallicity relative to solar
of [C/H]~$\sim$~$-2.8$ with no sign of redshift evolution over the
range $1.8\le z \le 4.1$ but a significant trend with over-densities 
\citep{Schaye03,Aguirre08}.
Given the expected low metallicities and the high ionization state of the gas, 
direct detection of metal absorption lines from underdense regions of the 
intergalactic medium (IGM) is beyond the scope of present day large telescopes. 
Statistical methods like pixel analysis 
are used instead \citep{Ellison00,Schaye03,Aracil04,Scannapieco06,Pieri10} and show that
metals must be present in the low-density regions. Even in regions where 
C~{\sc iv} absorption is detected directly, it is not clear however what is 
the main physical process that is maintaining the ionization state of the gas. 
In general, it is believed that photo-ionization keeps the gas ionized. 
However, it is probable that mechanical inputs from galactic winds can 
influence the ionization state of part of the IGM gas through collisional 
ionization at least in the proximity of galaxies. Therefore, it is important 
to simultaneously study different species covering a wide range of ionization 
states to get a better understanding of the metal enrichment and the different 
ionizing mechanisms at play. 

Recent hydrodynamical simulations \citep{Dave01,Fang01,Kang05,Bouche06,Bouche07} 
suggest that the missing baryons at low redshift, $z\sim 0-0.5$, and the missing 
metals at high redshift, $z\sim 2.5$, could reside in the warm-hot phase of the 
intergalactic medium (called WHIM) with $T \approx 10^{5} - 10^{7}$~K). 
Highly ionized species of oxygen such as O~{\sc vi}, O~{\sc vii} and O~{\sc viii} can be 
useful probes of the WHIM. While the strongest transitions of the latter two species 
have rest-wavelengths in the soft X-ray range, the spectral doublet 
O~{\sc vi}$\lambda\lambda$1032,1037 is seen in the near UV range and is therefore 
a useful probe of the gas at a temperature of $\sim$3$\times$10$^{5}$~K, temperature 
at which O$^{5+}$/O is maximum. 

It has been suggested that a large fraction of the conspicuous \os\ phase 
seen to be associated to high-$z$ damped Lyman-$\alpha$ systems may originate 
from collisionally ionized gas \citep{Fox07b}. However, photo-ionization can 
also maintain oxygen in a high ionization state and at relatively low 
temperature \citep[$T$~$\sim$ a few 10$^{4}$~K, see][]{Oppenheimer09}. 
Actually a large fraction of the O~{\sc vi} absorption seen at $z>2.5$ in 
quasars show Doppler broadening consistent with photo-ionization 
in the vicinity of the QSOs \citep[for example][]{Srianand00apm} but also in the diffuse 
intergalactic medium \citep[for example,][]{Bergeron02,Simcoe04}. 
While it is expected that the intrinsic ionizing spectrum of QSOs is hard enough
to maintain a high degree of ionization of oxygen in their vicinity, in the IGM, the
hardness of the ionizing spectrum will depend on the intrinsic spectral shape
of the ionizing sources and the IGM opacity at the H~{\sc i} and He~{\sc ii}
Lyman Limit \citep[][]{Haardt96, Fardal98}. 

An additional piece of information comes from QSO lines of sight transparent in 
the Lyman continuum [i.e a high-$z$ QSO line of sight without any
intervening Lyman limit system (LLS) blocking the
UV end of the spectrum]. It is then possible to observe the rest wavelength ranges of 
the H~{\sc i} and He~{\sc ii} Lyman-$\alpha$ forests and to compute the
ratio of the \he\ to the \hi\ optical depth (i.e the $\eta$ parameter). 
The bright QSO HE~2347$-$4342 at \zem = 2.885 \citep{Reimers97} is one such targets that  
attracted a lot of attention in the past years. It was shown that the \he\ opacity is 
``patchy'' in nature \citep[][]{Reimers97, Smette02}  and that $\eta$
decreases gradually from higher to lower redshift possibly due to a change
in the slope of the ionizing spectrum \citep[][]{Zheng04}. \citet{Shull04} 
discussed the small scale variations (over $\Delta z \approx 10^{-3}$) 
of $\eta$ and found an apparent correlation between high $\eta$ 
(less ionized He~{\sc ii}) and low \hi\ column density.  They ascribed 
these small scale $\eta$ variations to ``local ionization effects'' in the 
proximity of QSOs located close to the line of sight and having spectral 
indices ranging from $\alpha_{\rm s}$~=~0 to 3. 
\citet{Worseck07} reported the detection of 14 foreground QSOs in the field located 
close to the line of sight and could not find any convincing evidence for 
any transverse proximity effect from a decrease in the \hi\ absorption, although they 
did claim that the local UV spectrum inferred in the vicinity of three foreground 
QSOs appeared harder than expected, which is an indication of a transverse proximity effect. 
In turn these fluctuations could be due to an appreciable 
contribution of thermal broadening to the velocity width of absorption 
lines at high $N$(H~{\sc i}) \citep{Fechner07a}. 

In this paper and after a description of the observations (Section~2), 
we use a different approach involving Voigt profile fitting analysis of 
the H~{\sc i} and He~{\sc ii} absorption lines to measure $\eta$ (Section~3). 
We then report new detections of O~{\sc vi} absorption associated
with regions with low $\eta$ values (Section~4) and construct models 
of these regions (Section~5) before concluding in Section~6.

%========================== Observation ========================================================
\section{Observations}

The optical spectrum of HE~2347$-$4342 ($z_{\rm em}$~=~2.885) used in this 
study was obtained with the VLT UV Echelle Spectrograph (UVES) \citep{Dekker00} 
mounted on the ESO Kueyen 8.2-m telescope at the Paranal 
observatory in the course of the ESO-VLT large programme `The Cosmic 
Evolution of the IGM' \citep{Bergeron04}. HE~2347$-$4342 was 
observed through a 1-arcsec slit (with a typical seeing of 0.8 arcsec) 
for 12~h with central wavelengths adjusted to 3460 and 5800~\AA~ in the 
blue and red arms, respectively, using dichroic \#1 and for 
another 14~h with central wavelengths at 4370 and 8600\AA~ in the blue 
and red arms with dichroic \#2. The raw data were reduced using the latest 
version of the UVES pipeline \citep{Ballester00} which is 
available as a dedicated context of the MIDAS data reduction software. 
The main function of the pipeline is to perform a precise inter-order 
background subtraction for science frames and master flat fields, 
and to apply an optimal extraction to retrieve the object signal, 
rejecting cosmic ray impacts and performing sky subtraction at the same 
time. The reduction is checked step-by-step. Wavelengths 
are corrected to vacuum-heliocentric values and individual one-dimensional 
spectra are combined. Air-vacuum conversions and heliocentric 
corrections were done using standard conversion equations 
\citep{Edlen66,Stumpff80}. Addition of individual exposures is
performed by adjusting the flux in individual exposures to the same 
level and inverse variance weighting the signal in each pixel. 
Great care was taken in computing the error spectrum while combining 
the individual exposures. Our final error in each pixel
is the quadratic sum of the weighted mean of errors in the different 
spectra and the scatter in the individual flux measurements. 
Errors in individual pixels obtained by this method are consistent
with the rms dispersion around the best fitted continuum in regions free of
absorption lines. The final combined spectrum covers the wavelength range of 
3000 to $10,000$~{\AA}. A typical SNR$\sim$60 per pixel (of 0.035~\AA) 
is achieved over the whole wavelength range of interest for a spectral resolution 
of $R \sim 45,000$. The detailed quantitative description of data 
calibration is presented in \citet{Aracil04} and \citet{Chand04}. 

We use the continuum normalized FUSE data provided by Dr. Zheng. The details 
of the data reduction and continuum normalization can be found in 
\citet{Zheng04} \footnote { We have obtained individual spectra reduced using 
Calfuse 3.2.1 version from http://fuse.iap.fr/interface.php. We combine LIF spectra after 
correcting for the background by demanding zero flux in the core of strong 
saturated absorptions in the wavelength range 1130--1185 \AA. When we follow 
the same continuum fitting and re-binning procedures, we find the new data follow 
the structures (both in wavelength and flux) as seen in the data of \citet{Zheng04} 
very well and fitting results are not changed. So, results presented in this paper 
will not change when one uses the new pipeline for the data reduction. 
Whereas this work was already completed, new COS data on this object
were reported by \citet{Shull10}. As the COS spectrum is found 
to be consistent (see their Fig. 3) with the FUSE spectra used here, 
this has no consequence on the results of this paper.}. 
The original data have typical resolution of $R=20,000$ 
and signal-to-noise ratio $\sim$5 in the long wavelength range 
($\lambda$~$>$~1050~\AA). Following \citet{Zheng04}, we have re-binned 
this data to 0.1~\AA,\  which leads to an effective resolution 
of $R\sim4000$. We restrict ourselves to the wavelength range with 
SNR~$>$~10. This corresponds to a redshift range 2.58$\le$ z$\le$2.70 
or a velocity range of $\sim$$10,000$~\kms around a central 
redshift of $z=2.6346$ (see Fig.~\ref{total_vplot}).
%

%========================  Section :: HeII to HI ratio =========================================

\section{$N$(\he)/$N$(\hi) ratio}

%____________________ Total velocity plot ______________________________________________________
\begin{figure*} {}{}{}{}{}{}
\includegraphics[width=0.7\textwidth, angle = 270]{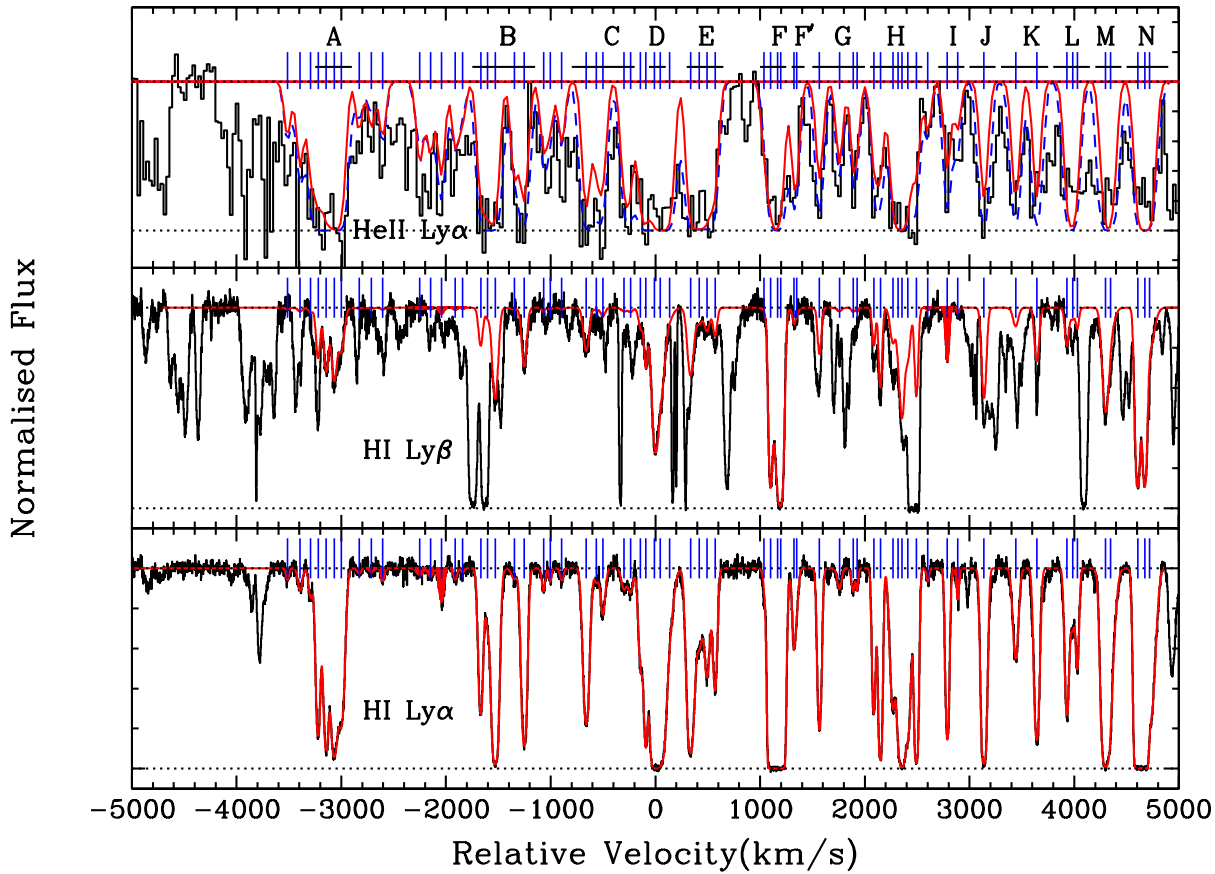}
\caption{H~{\sc i} and He~{\sc ii} Lyman-$\alpha$ and H~{\sc i} Lyman-$\beta$ absorption profiles 
on a velocity scale with origin $z=2.6346$. Different velocity ranges where $\eta$ is measured 
through $\chi^2$ minimization are indicated with horizontal lines and labelled by letters in 
alphabetical order from left to right. Vertical tick marks located above the absorption 
profiles show the positions of the individual Voigt profile components used to fit the 
Lyman-$\alpha$ and Lyman-$\beta$ H~{\sc i} lines together. The best fit models for H~{\sc i} 
Lyman-$\beta$ and Lyman-$\alpha$ are overplotted to the data. The FUSE spectrum is shown in the 
top panel together with the best fitted profile obtained from scaling the H~{\sc i} column 
densities by  the fitted $\eta$ parameter and assuming thermal (solid red line) or turbulent 
(dashed blue line) broadening for the He~{\sc ii} lines.}
\label{total_vplot}
\end{figure*}
%___________________________ Chi^2 Curves _______________________________________________________

\begin{figure*}
\centerline{
\vbox{
\hbox{
\includegraphics[height=5.7cm,width=4.5cm,angle= 0]{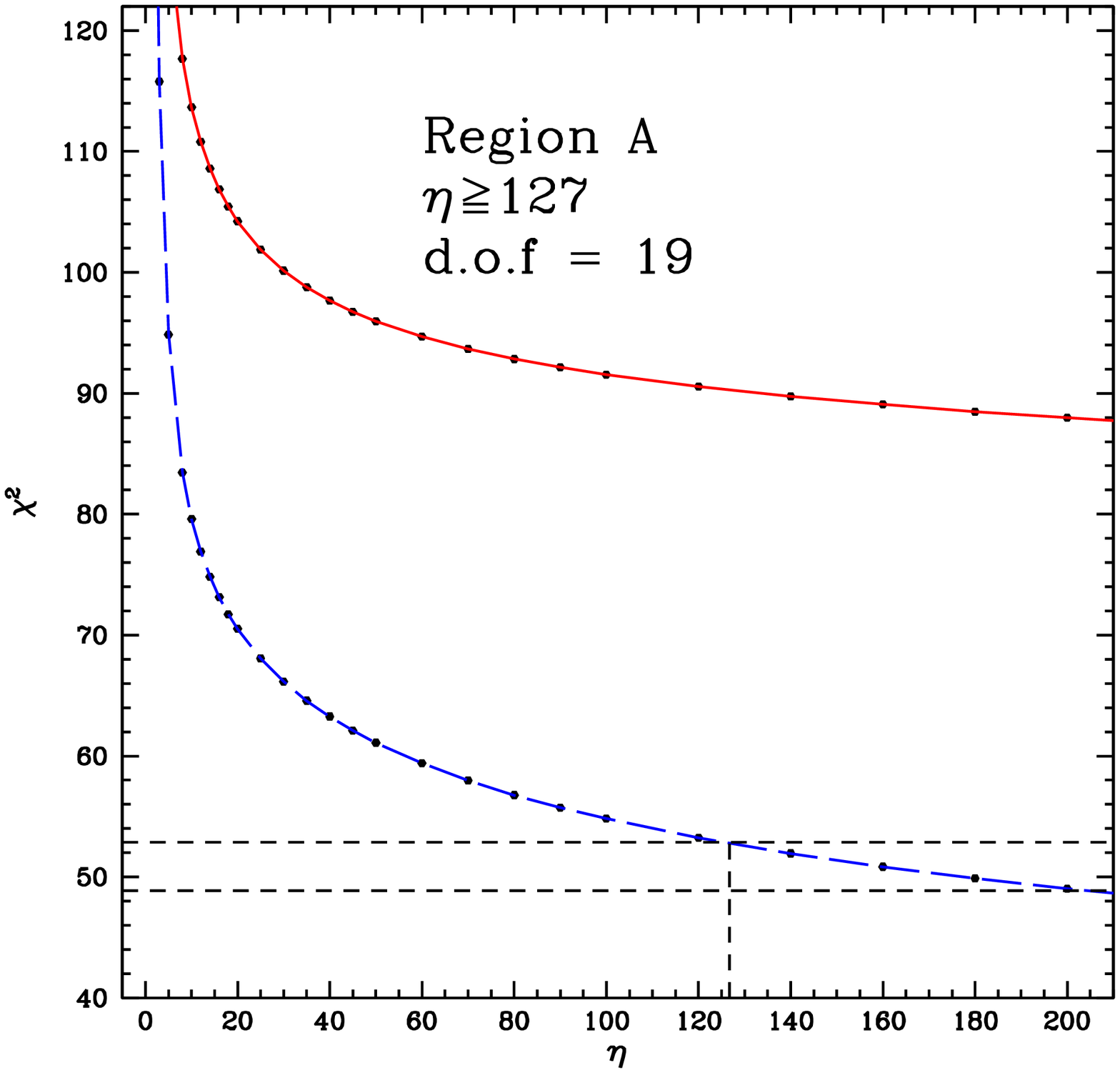}
\includegraphics[height=5.7cm,width=4.5cm,angle= 0]{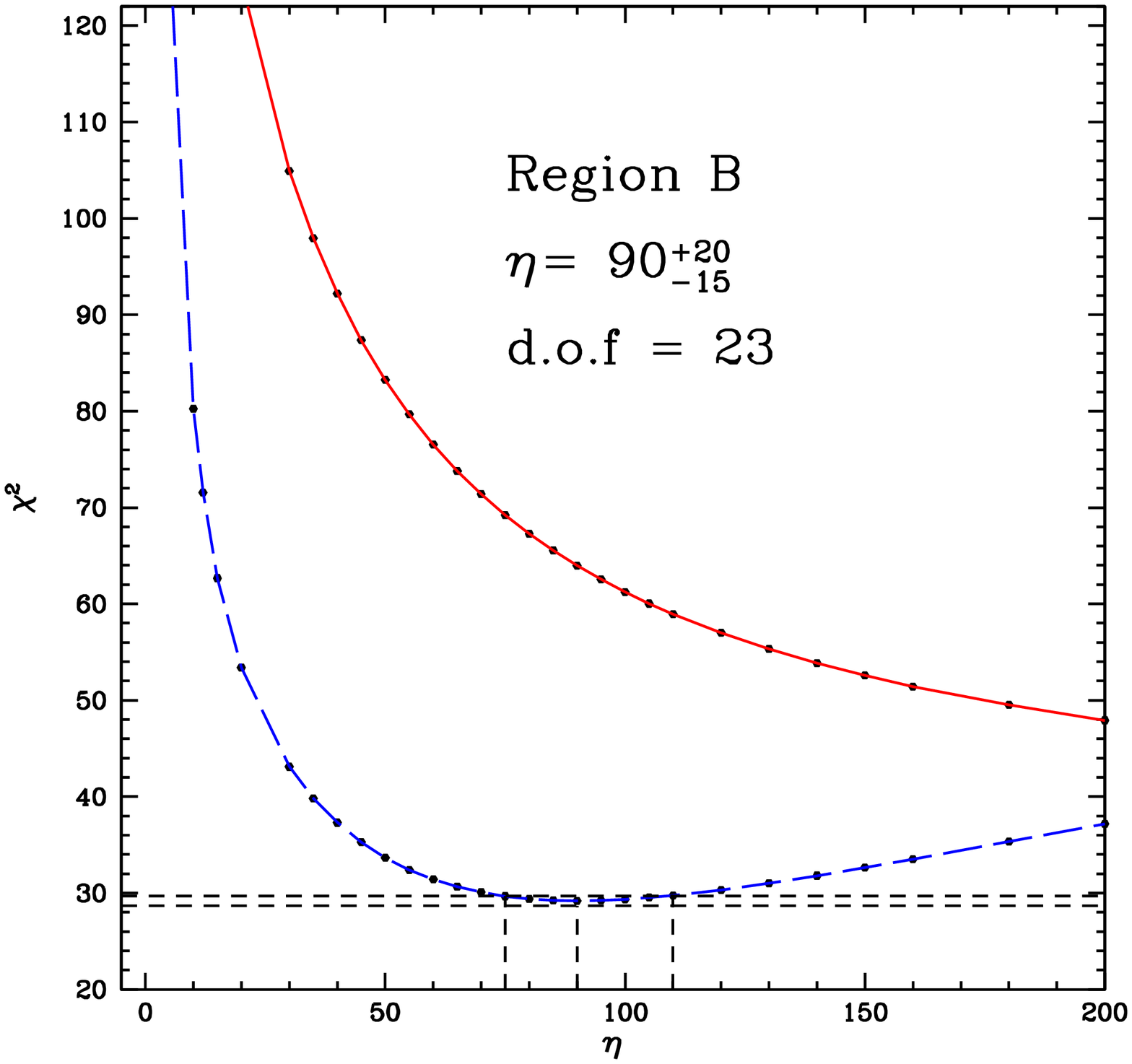}
\includegraphics[height=5.7cm,width=4.5cm,angle= 0]{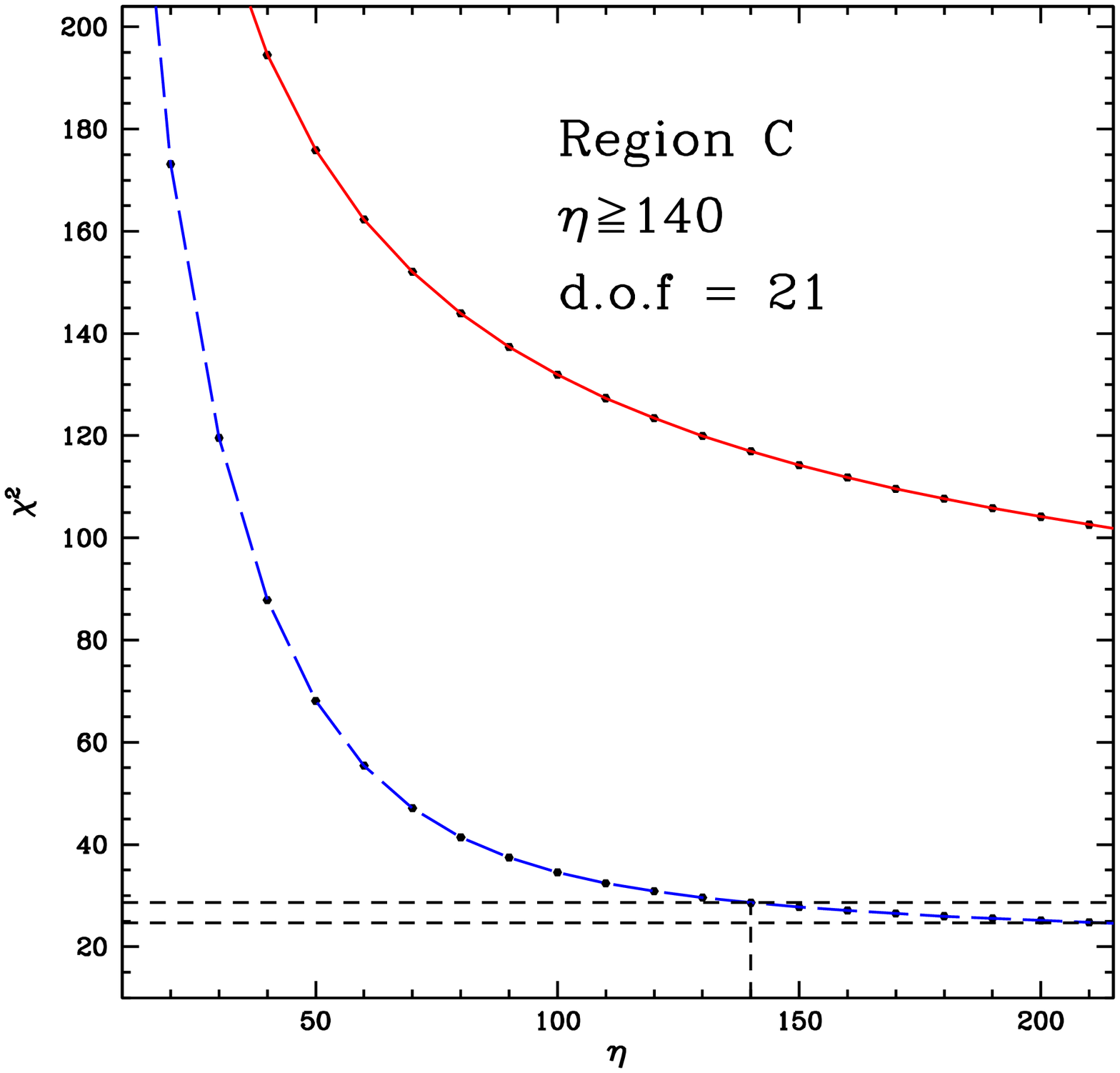}
\includegraphics[height=5.7cm,width=4.5cm,angle= 0]{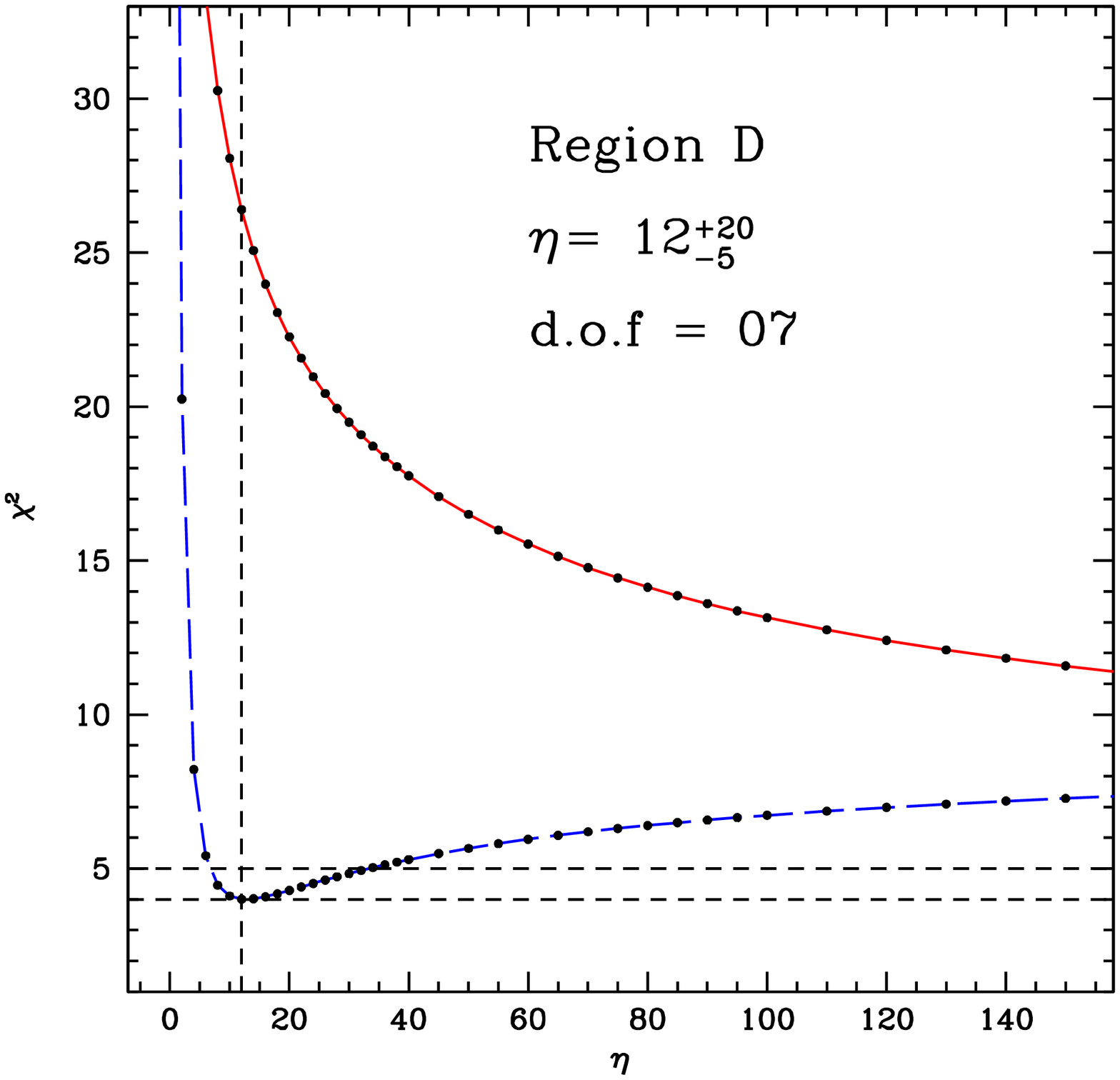}
}
\hbox{
\includegraphics[height=5.7cm,width=4.5cm,angle= 0]{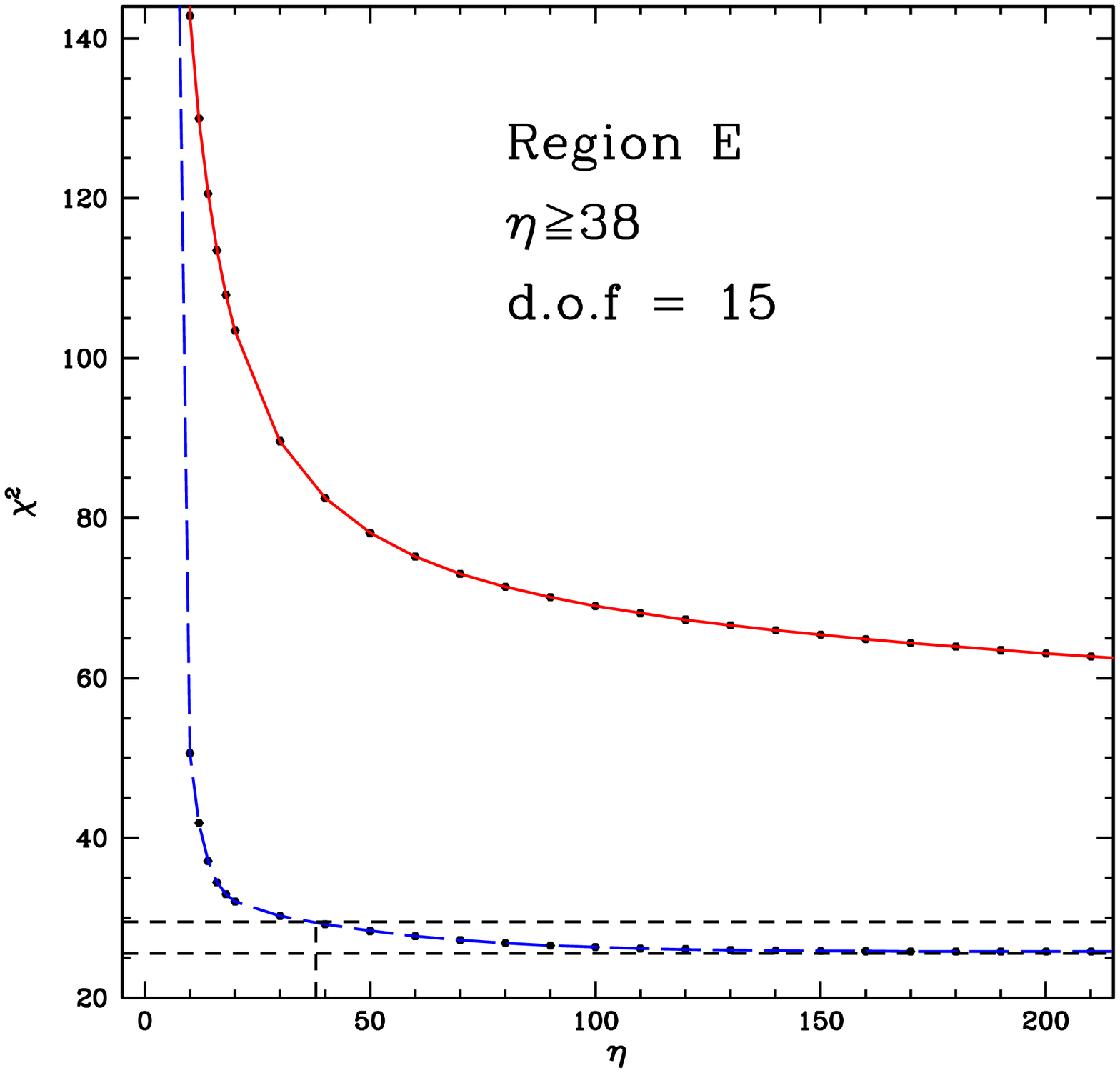}
\includegraphics[height=5.7cm,width=4.5cm,angle= 0]{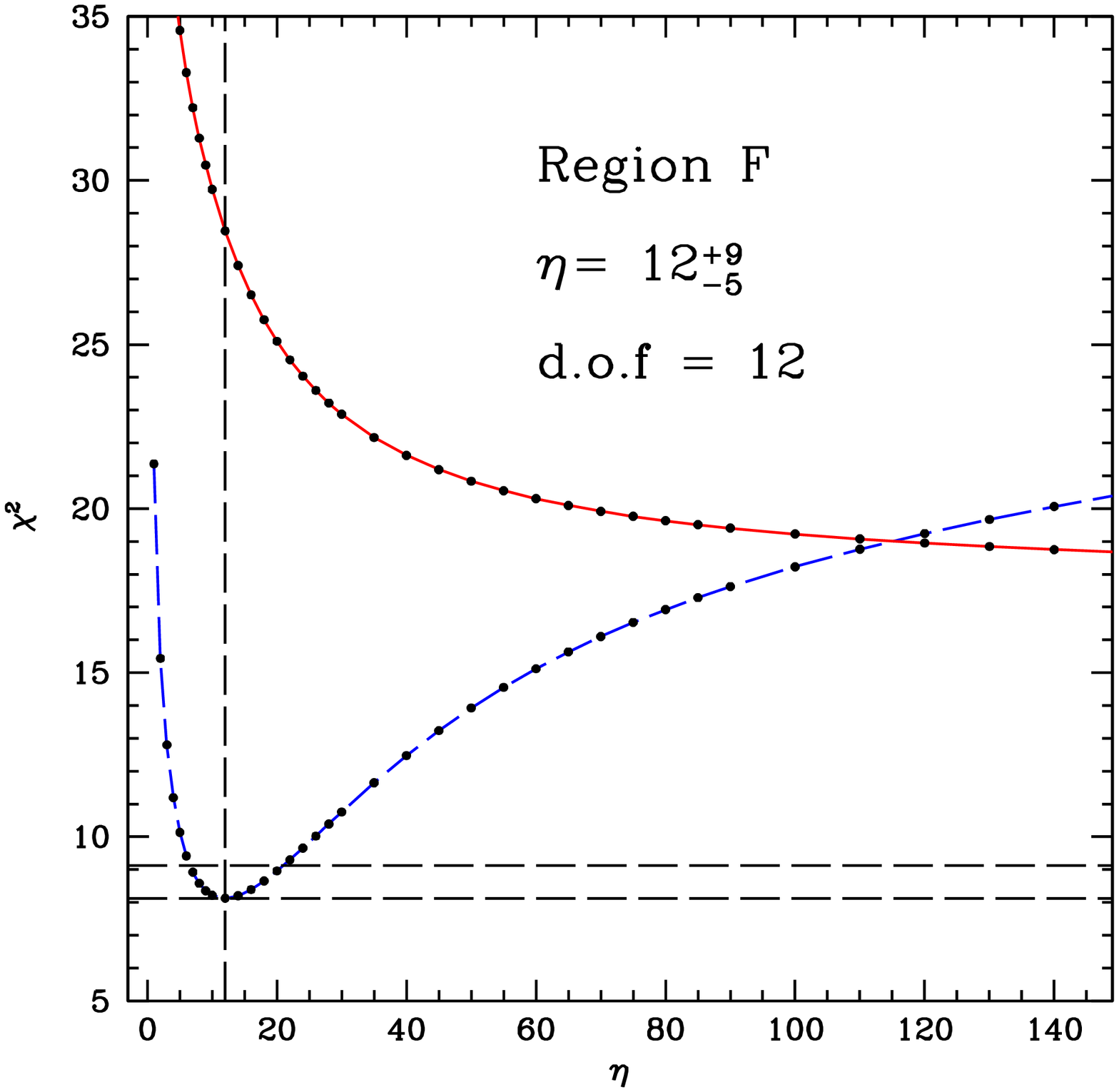}
\includegraphics[height=5.7cm,width=4.5cm,angle= 0]{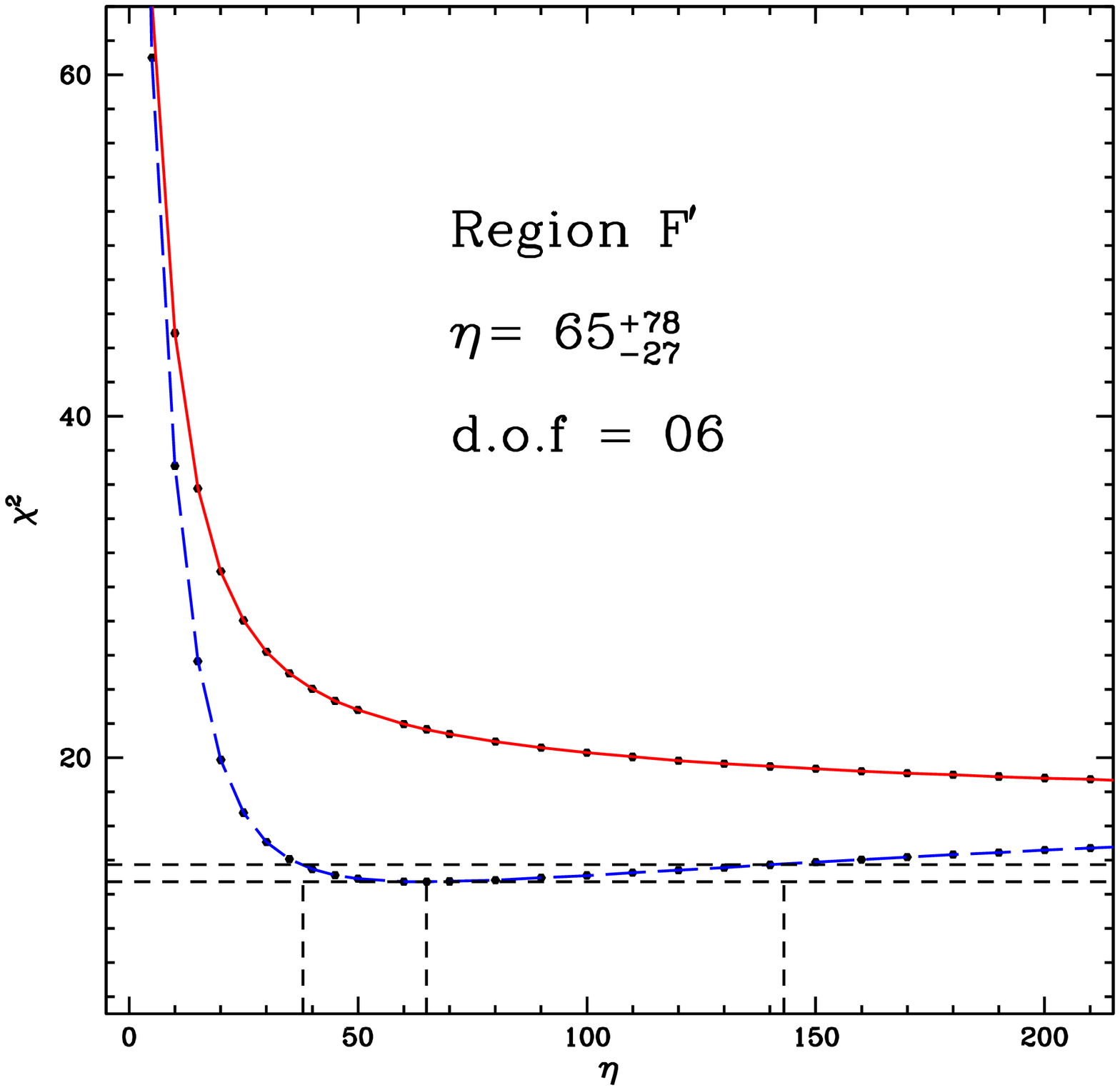}
\includegraphics[height=5.7cm,width=4.5cm,angle= 0]{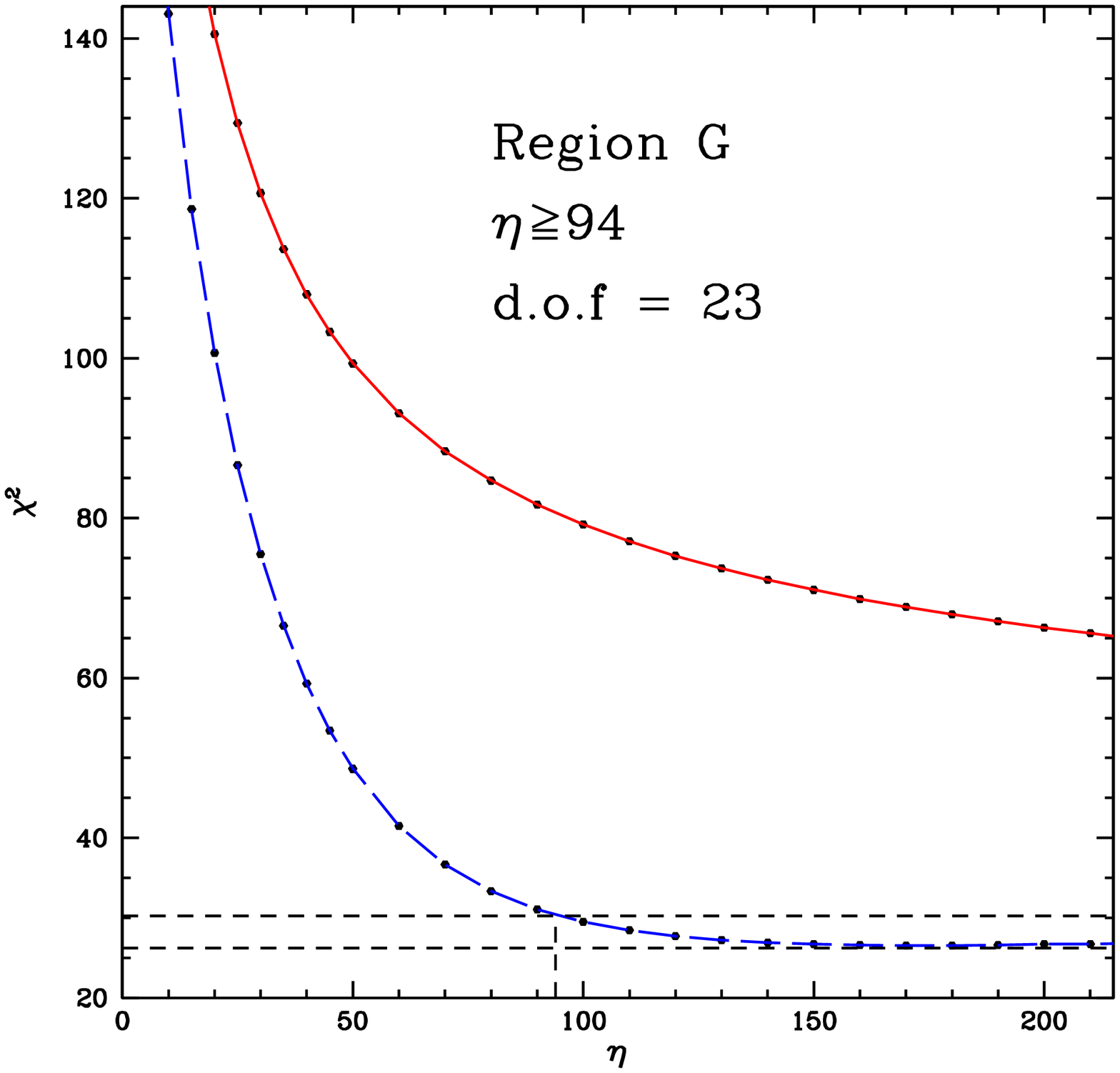}
}
\hbox{
\includegraphics[height=5.7cm,width=4.5cm,angle= 0]{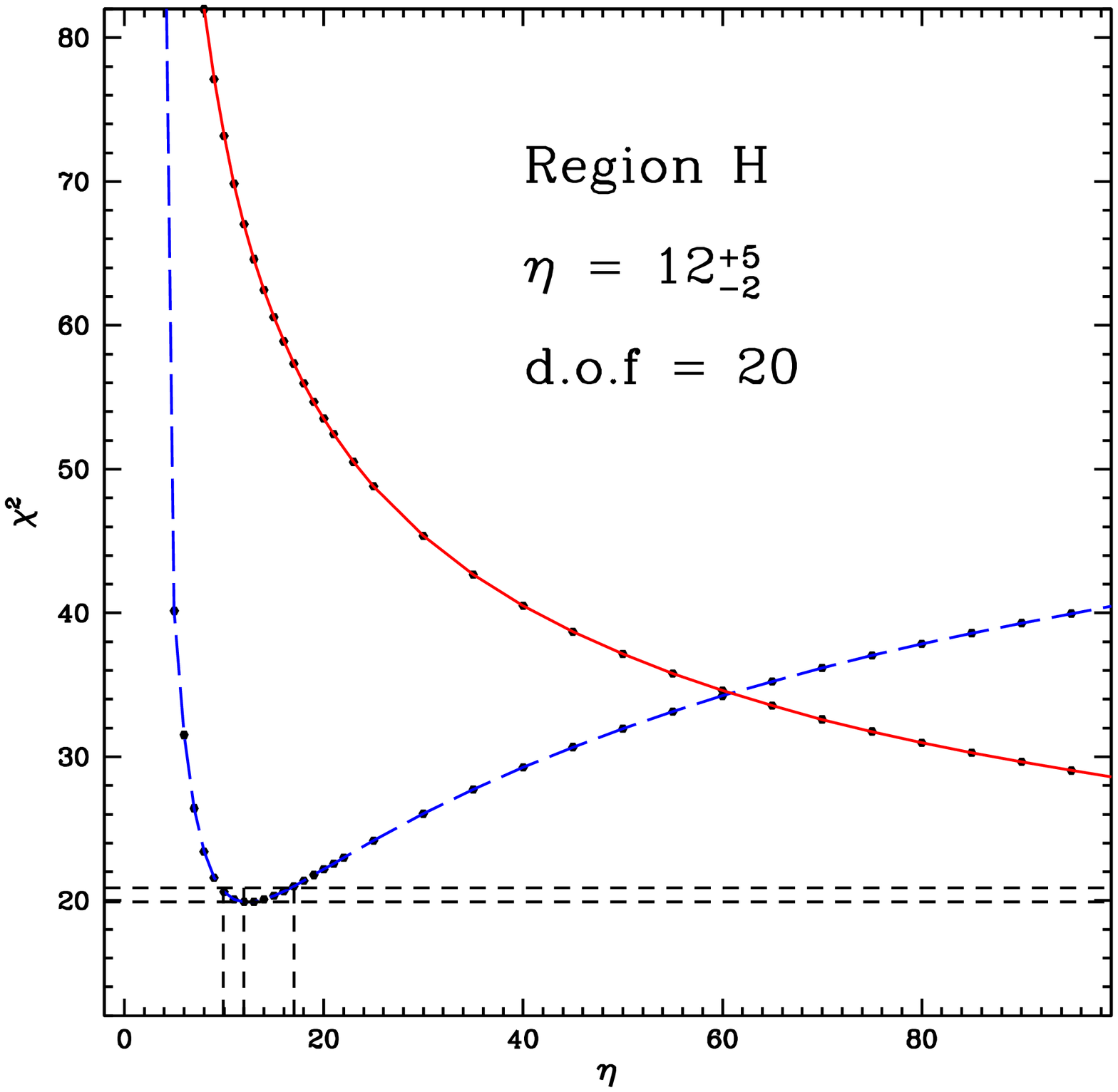}
\includegraphics[height=5.7cm,width=4.5cm,angle= 0]{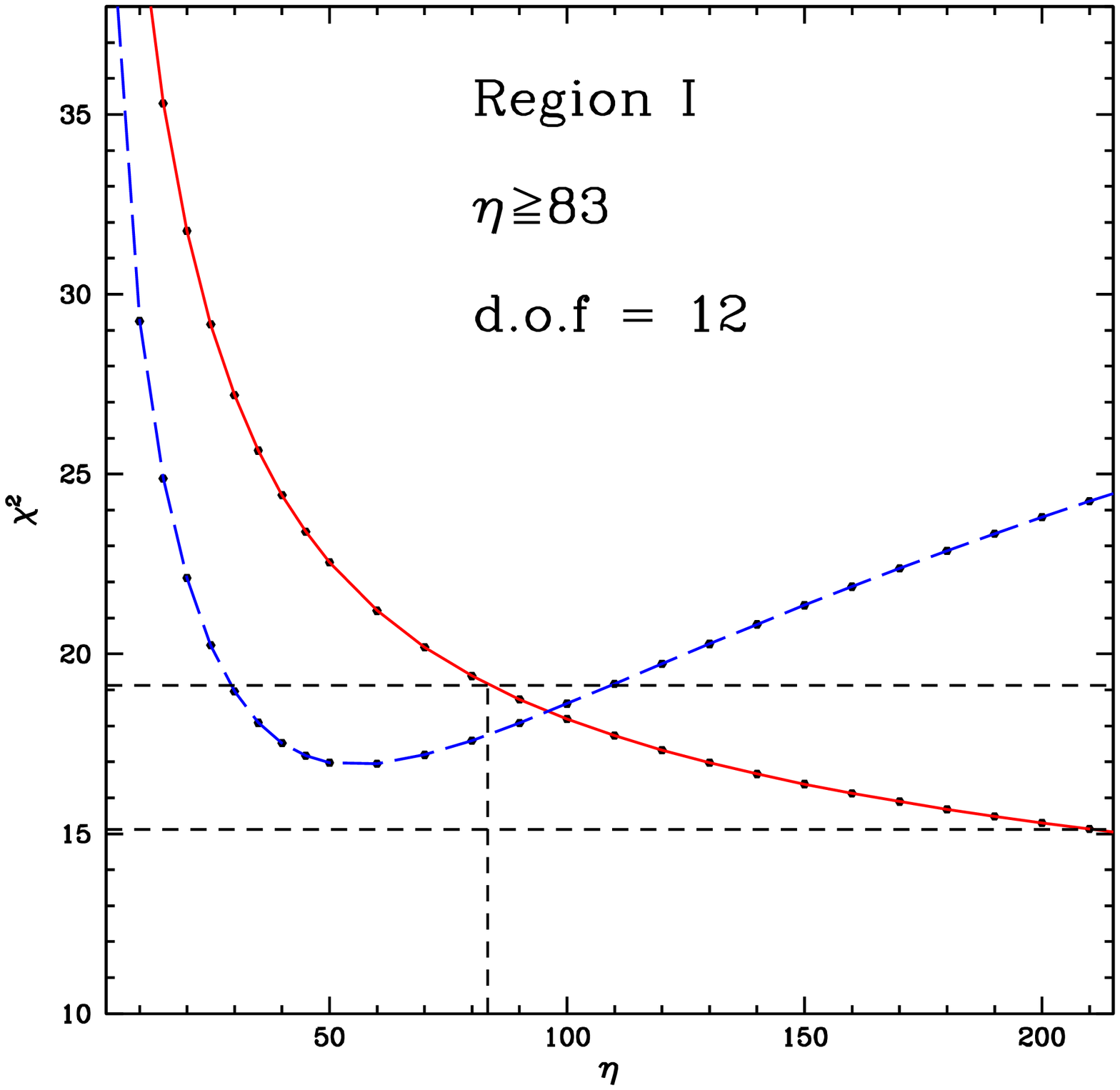}
\includegraphics[height=5.7cm,width=4.5cm,angle= 0]{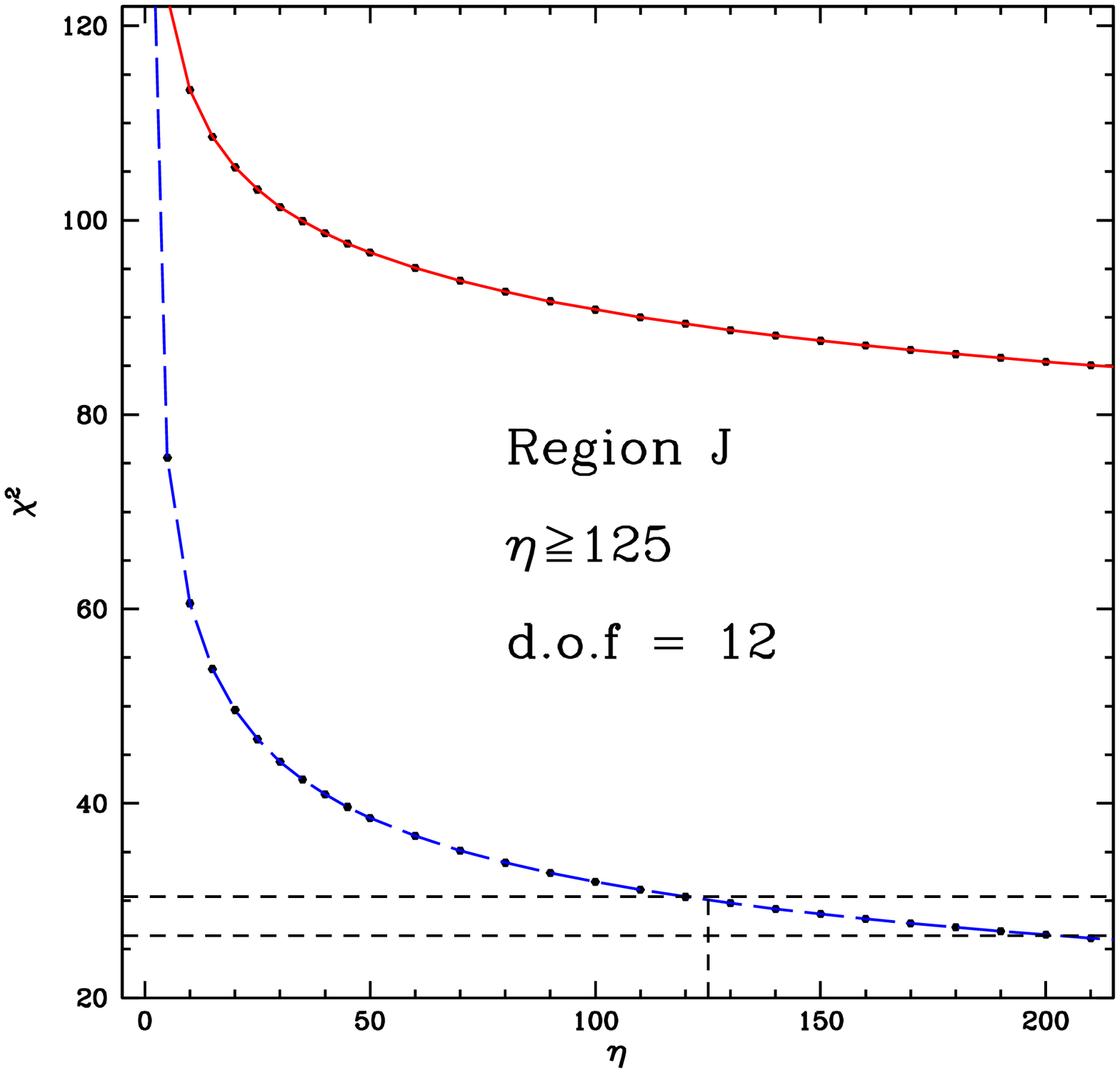}
\includegraphics[height=5.7cm,width=4.5cm,angle= 0]{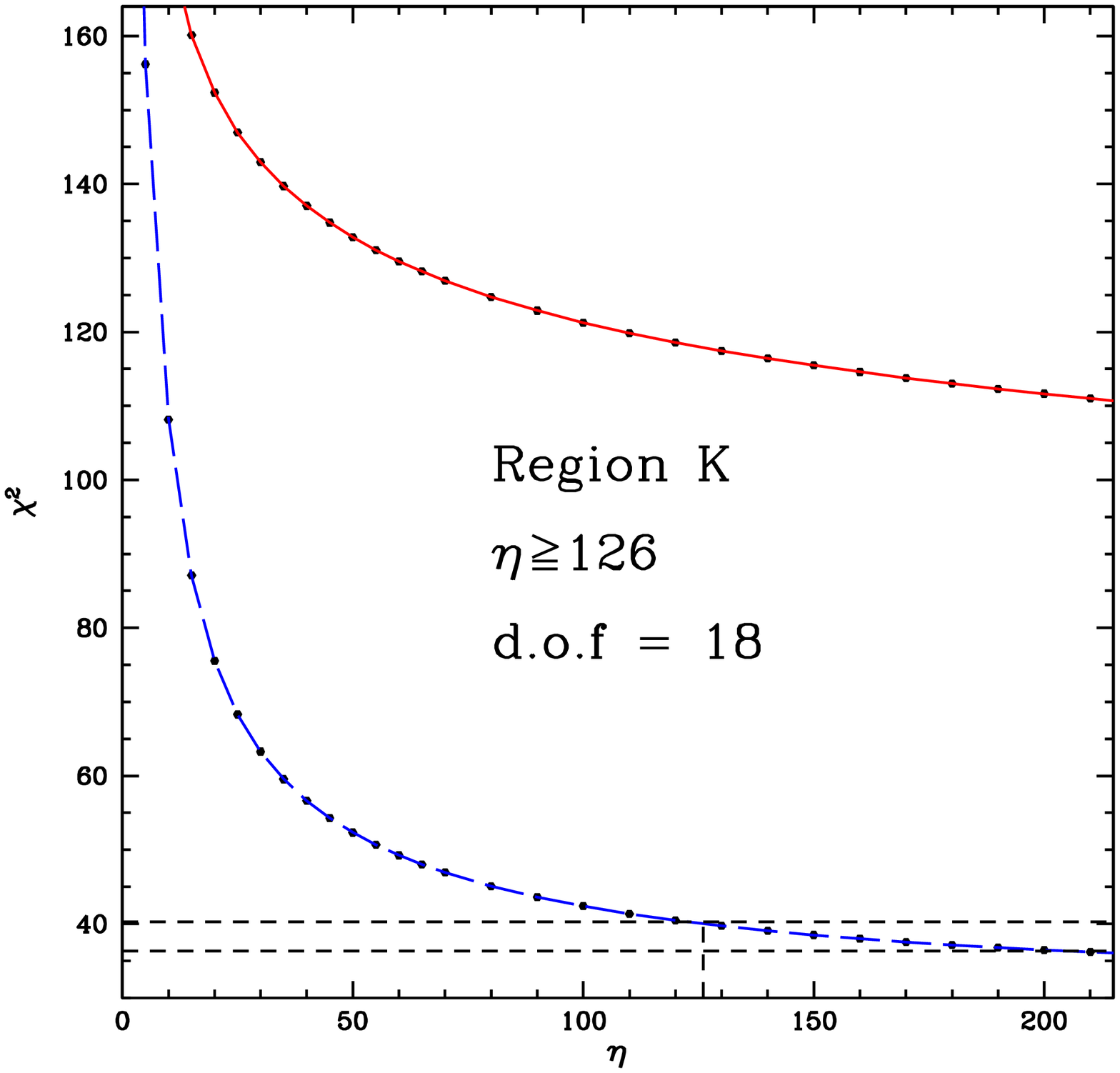}
}
\hbox{
\includegraphics[height=5.7cm,width=4.5cm,angle= 0]{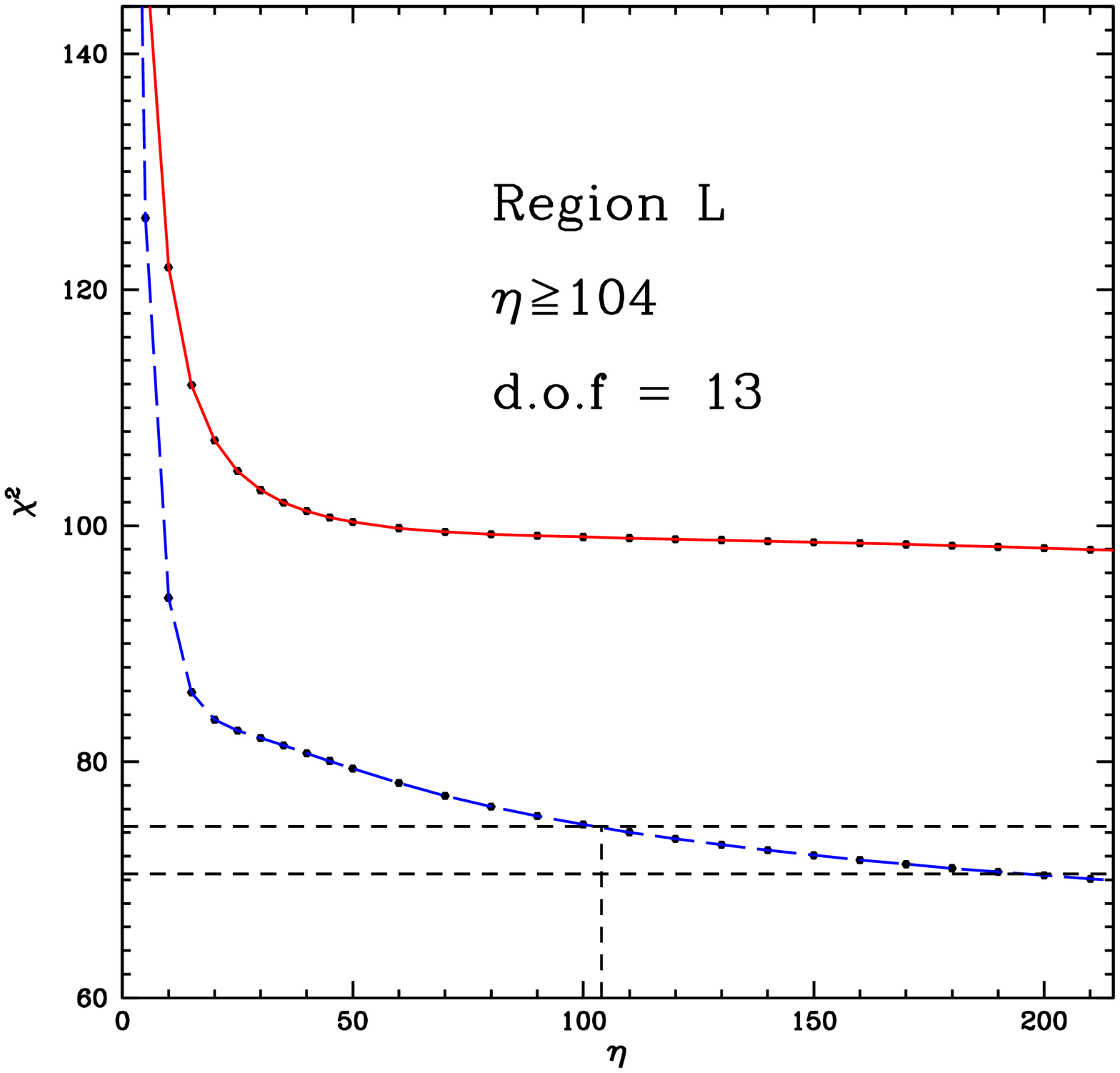}
\includegraphics[height=5.7cm,width=4.5cm,angle= 0]{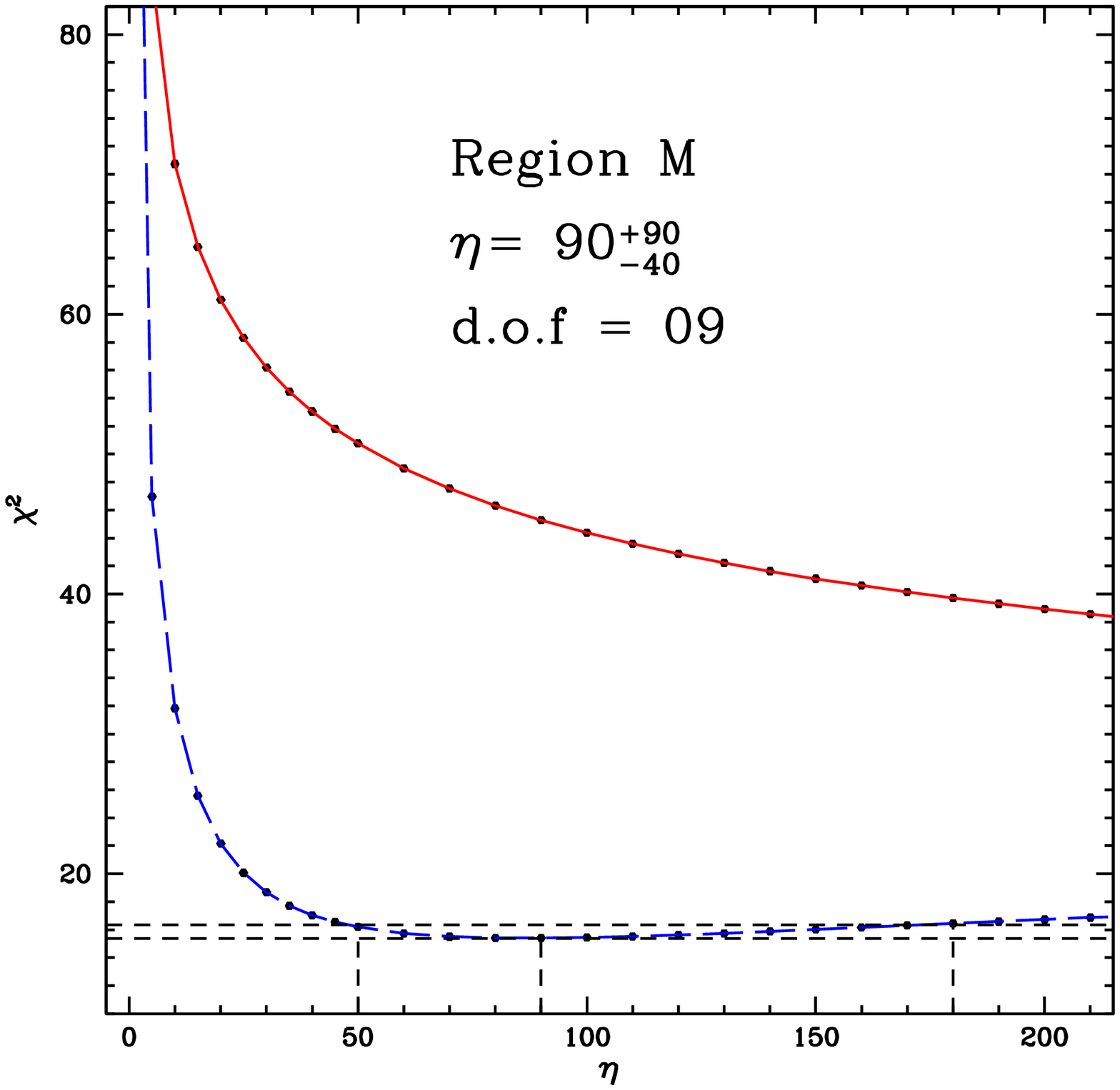}
\includegraphics[height=5.7cm,width=4.5cm,angle= 0]{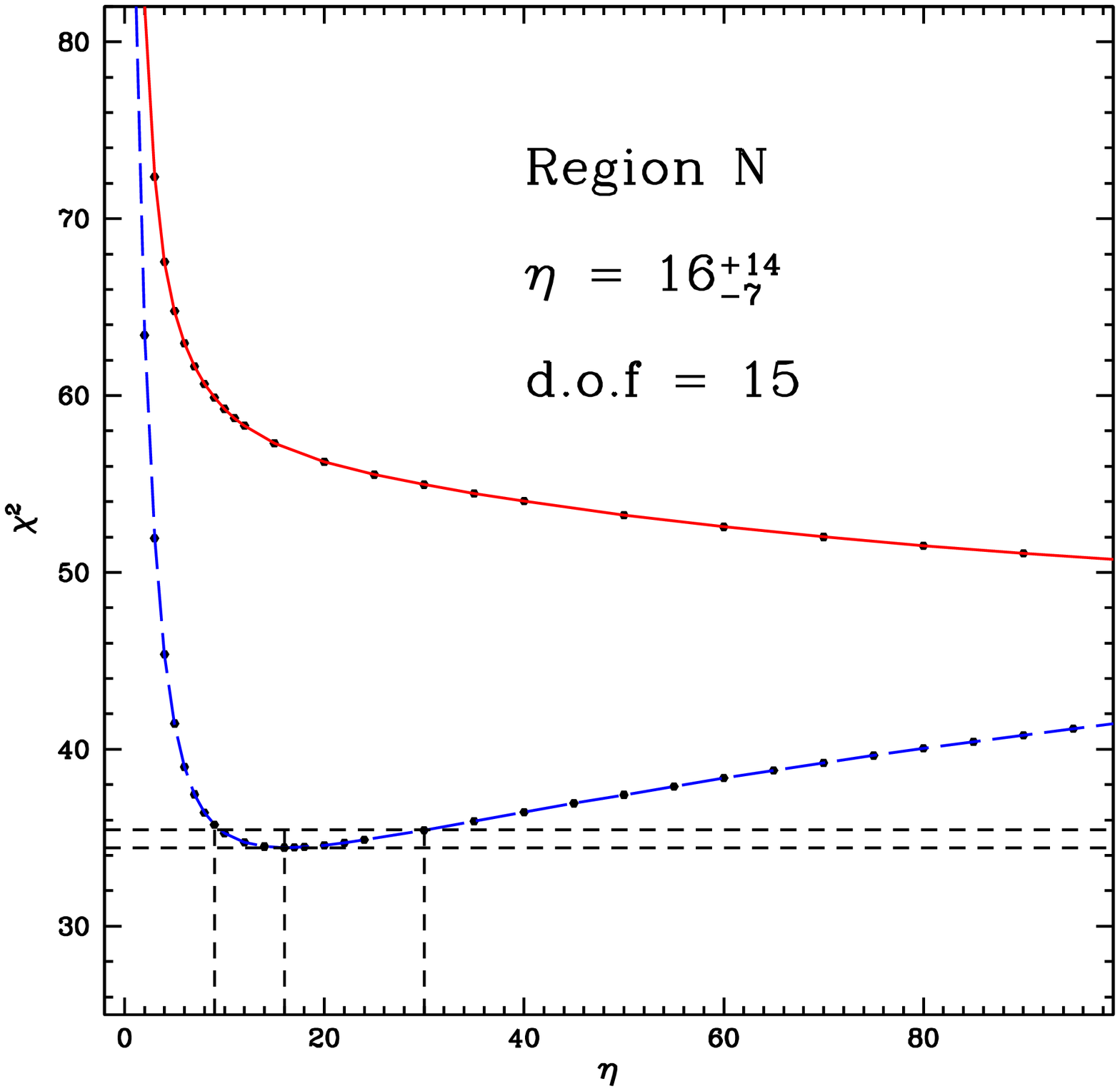}
}
}
}
\caption{$\chi^2$ resulting from the comparison of the He~{\sc ii} absorption with a model 
profile obtained by scaling the H~{\sc i} column density with the parameter $\eta$ as a 
function of $\eta$ for the different regions defined in Fig.~\ref{total_vplot}. The dashed 
blue and solid red curves are for the two extreme cases, respectively: 
$b$(He~{\sc ii})~=~$b$(H~{\sc i}) (i.e turbulent broadening) or 
$b$(He~{\sc ii})~=~0.5$\times$$b$(H~{\sc i}) (i.e pure thermal broadening).}
\label{chisq_plot}
\end{figure*}
%_________________________________________________________________________________

In this Section we concentrate on the column density ratio 
$\eta$~=~$N$(He~{\sc ii})/$N$(H~{\sc i}) over the redshift range 2.58$\le z\le 2.70$ where the 
FUSE data show relatively good signal to noise ratio. This range roughly corresponds to a 
relative velocity range of $-$4000 \kms\ to $+$ 5000 \kms\ around the strong \os\ absorber 
seen at \zabs~=~2.6346 (see Fig.~\ref{total_vplot}).

As a first step we fitted simultaneously the Lyman-$\alpha$ to Lyman-$\gamma$ profiles 
when possible, e.g. when the Lyman-$\beta$ and/or Lyman-$\gamma$ lines are not 
blended with another intervening Lyman-$\alpha$ line. Then we compare the He~{\sc ii} 
absorption profile with a model with the same components as the H~{\sc i} model, scaling 
the fitted H~{\sc i} column densities by the parameter $\eta$. We consider two alternatives: 
In the first case we use the same Doppler parameter for H~{\sc i} and He~{\sc ii} 
(assuming turbulent broadening); in the second case we give the He~{\sc ii} $b$-parameter 
the value expected from thermal broadening (i.e $b$(He~{\sc ii})~=~0.5$\times$$b$(H~{\sc i})). 
The best fitted values of $\eta$ is obtained by $\chi^2$ minimization. While fitting the 
He~{\sc ii} profiles we use a Gaussian convolving function to correctly represent the FUSE 
spectral resolution. For Voigt-profile decomposition we have used the fitting code developed
by \citet{Khare97}. 

As the FUSE data are of much lower resolution and SNR than the UVES data, we cannot estimate 
$\eta$ for individual H~{\sc i} components. Instead, we have singled out 15 small regions named 
as A, B, C etc., in Fig.~\ref{total_vplot} and we derive the best $\eta$ value over each region. 
We wish to point out that the approach we have taken here is very different from previous studies. 
Indeed, \citet{Shull04} used apparent optical depth in Lyman-$\alpha$ only (AOD) method, 
whereas \citet{Kriss01} and \citet{Zheng04} used Gaussian decomposition and \citet{Fechner07a} 
scale the whole \hi\ spectrum by $\eta$~=~$4\times$$\tau_{\rm HeII}/\tau_{\rm HI}$ to fit 
the \he\ data. In all these studies only H~{\sc i} Lyman-$\alpha$ is used 
\footnote {limited amount of analysis of Lyman-$\beta$ have been done by \citet{Zheng04}.}. 
This is the use of the H~{\sc i} optical depths in all available Lyman series lines 
that allows us to discriminate between thermal and turbulent broadening. 
%
%_____________________________ log eta Vs. log NHI ___________________________________
\begin{figure}{}{}{}{}{}{}
\includegraphics[height=9.0cm,width=8.0cm,angle= 0]{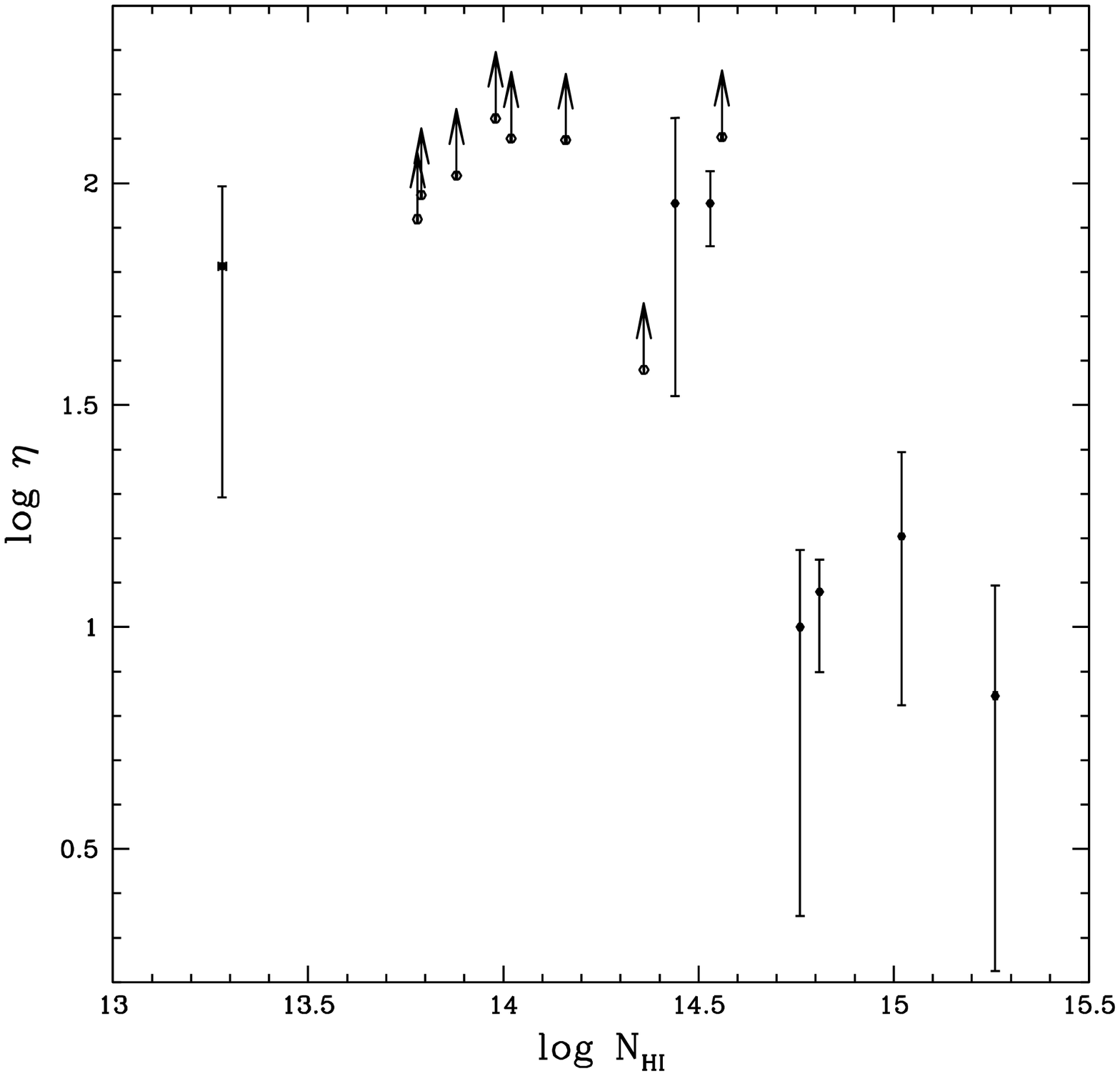}
\label{lgeta_lgNH}
\caption{The $\eta$ values for the 15 regions singled out in Fig.~\ref{total_vplot} 
         are plotted against the neutral hydrogen column density. }
\end{figure}

%_______________________________________________________________________________________

The best fitted Voigt profiles to the Lyman-$\alpha$ and Lyman-$\beta$ absorption lines are 
shown in the bottom and middle panels of Fig.~\ref{total_vplot}. The top panel shows the 
best fitted He~{\sc ii} Lyman-$\alpha$ line with the two assumptions on the Doppler parameter 
discussed above. The $\chi^2$\ curves as a function of $\eta$\ for the different regions singled 
out in Fig.~\ref{total_vplot} are shown in Fig.~\ref{chisq_plot}. The solid and dashed lines in 
these plots represent the cases of thermal and turbulent broadening respectively. In most cases 
the \chisq\ curve shows a clear minimum thereby allowing us to discriminate between the turbulent 
and thermal cases, and to derive the best fitted value of $\eta$. Errors are estimated from the 
range of $\eta$ values corresponding to $\Delta \chi^2$~=~$\pm$1 around the minimum. There are 
regions, especially when the He~{\sc ii} Lyman-$\alpha$\ line is saturated, for which the 
$\chi^2$ curve flattens (e.g. regions {\bf E} and {\bf G}), we have only one-sided limit. 
In these cases we define the 2$\sigma$ lower limit of $\eta$\ as the value corresponding to a 
$\chi^2$ equal to $\chi^2$ of the flat part of the curve plus four. The shapes of the $\chi^2$\ 
curves are not symmetric which is a natural consequence of line saturation.

It is clear from the Figure that, apart from region {\bf I}, the $\chi^{2}$ values
are smaller in case of turbulent broadening and that minima are reached only in that case. 
In the case of thermal broadening, the $\chi^2$ curves seem to saturate to some asymptotic 
value probably because the observed He~{\sc ii} profiles are too broad to be reproduced by 
the model. Thus the exercise presented here shows that the width of He~{\sc ii} 
Lyman-$\alpha$ lines are consistent with the $b$-parameter derived from H~{\sc i} lines.

If the gas is optically thin and photo-ionized by a UV background dominated by QSOs, we would 
expect $\eta$ to be in the range 40$-$400 depending on the exact spectral index and the IGM opacity. 
In the case of self-shielded optically thick gas, $\eta$ could be even higher \citep{Fardal98}. 
Four regions ({\bf D}, {\bf F}, {\bf H} and {\bf N}) in Fig.~\ref{chisq_plot} have $\eta\le40$.
These regions are associated with large H~{\sc i} column densities as can be seen on 
Fig.~\ref{lgeta_lgNH} where log~$\eta$ is plotted against log~$N$(H~{\sc i}) as measured in the 
different regions. This correlation was already noted in earlier works. Fechner \& Reimers (2007) 
argued that this can be explained if the thermal broadening of lines are also important.

In the following, we will use additional information on metal lines observed in the UVES
spectrum to discuss further the ionization state of the gas in these regions.

%===================== Section :: Individual Systems ===================================

\section{Regions with low $\eta$\ values} 

In the previous Section, we have shown that the $N$(He~{\sc ii})/$N$(H~{\sc i}) ratio can
be explained over most of the observed spectrum by ionization of the gas by the UV background
except in four regions: {\bf D} (\zabs = 2.6346), {\bf F} (\zabs = 2.6498),  {\bf H} 
(\zabs = 2.6624) and  {\bf N} (\zabs = 2.6910). The presence of \os\ and \cf\ absorption in 
systems showing low values of $\eta$ may yield interesting clues about (i) the nature of the 
ionizing radiation, (ii) the effect of thermal/turbulent broadening and (iii) the possible 
mechanical feedback from winds.

\begin{figure*}{}{}{}{}{}{}
\centerline{
\vbox{
\hbox{
\includegraphics[height= 9.0cm,width= 8.0cm,angle= 0]{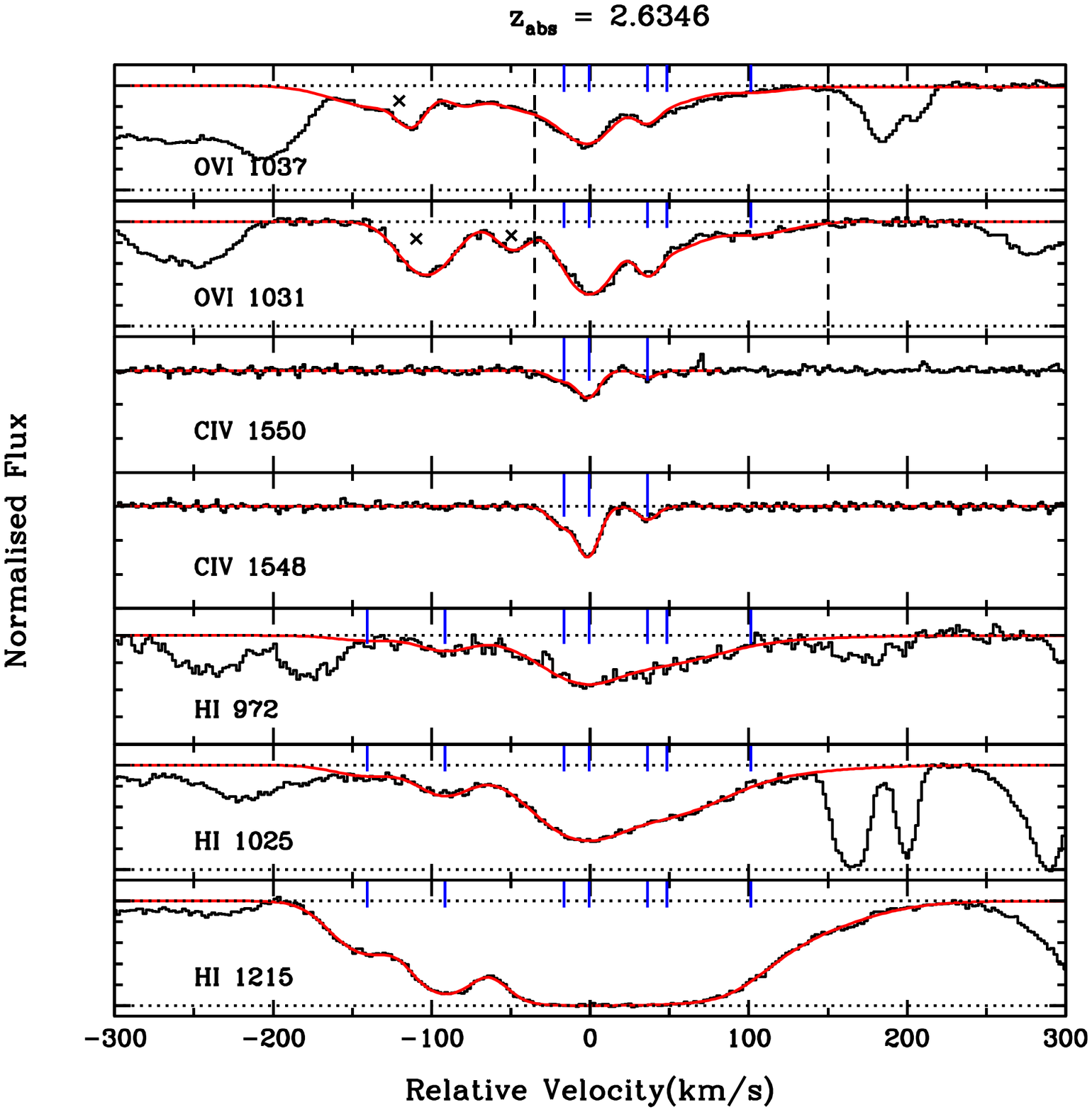}
\includegraphics[height= 9.0cm,width= 8.0cm,angle= 0]{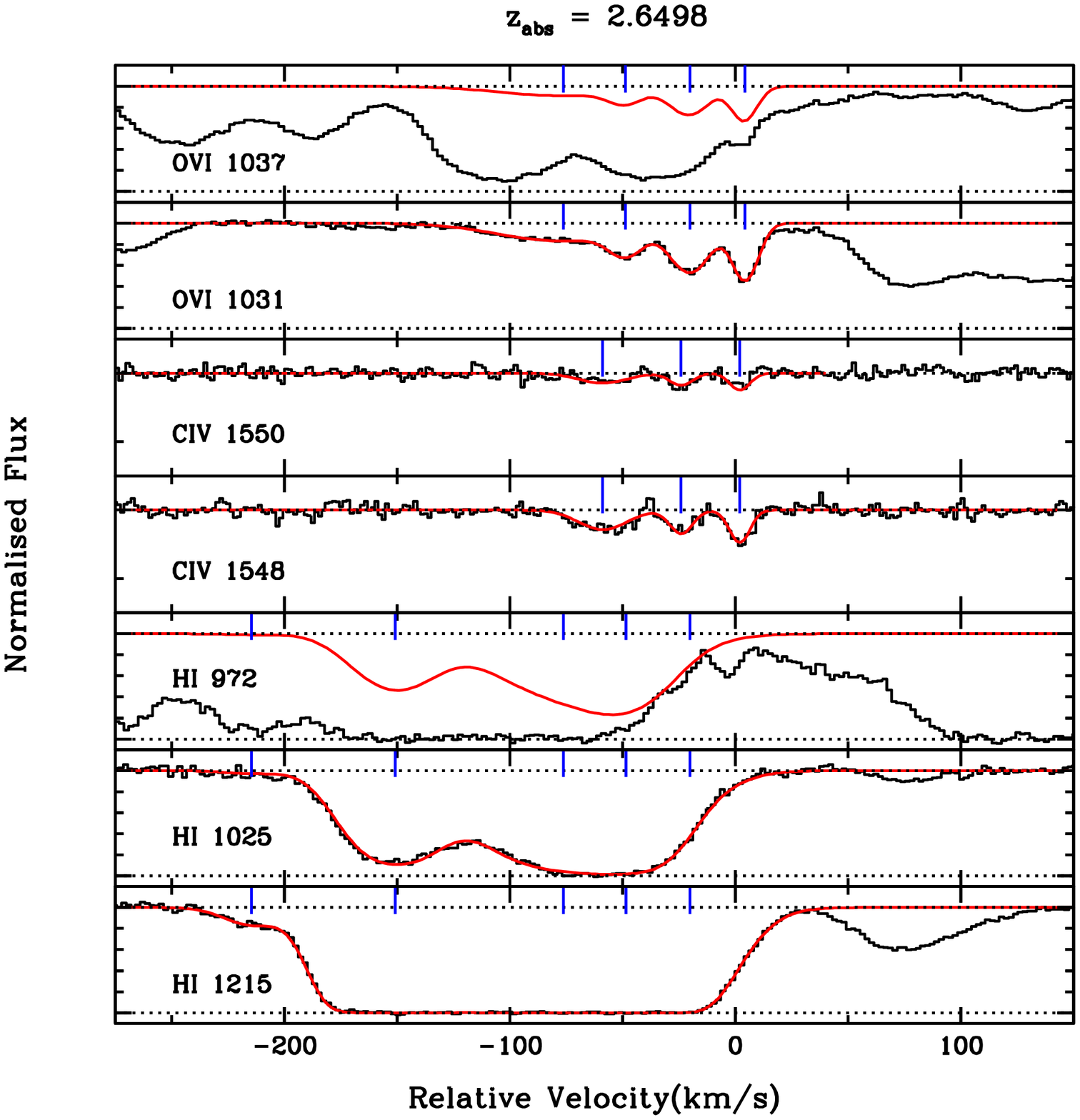}}
\hbox{
\includegraphics[height= 8.0cm,width= 8.cm,angle= 0]{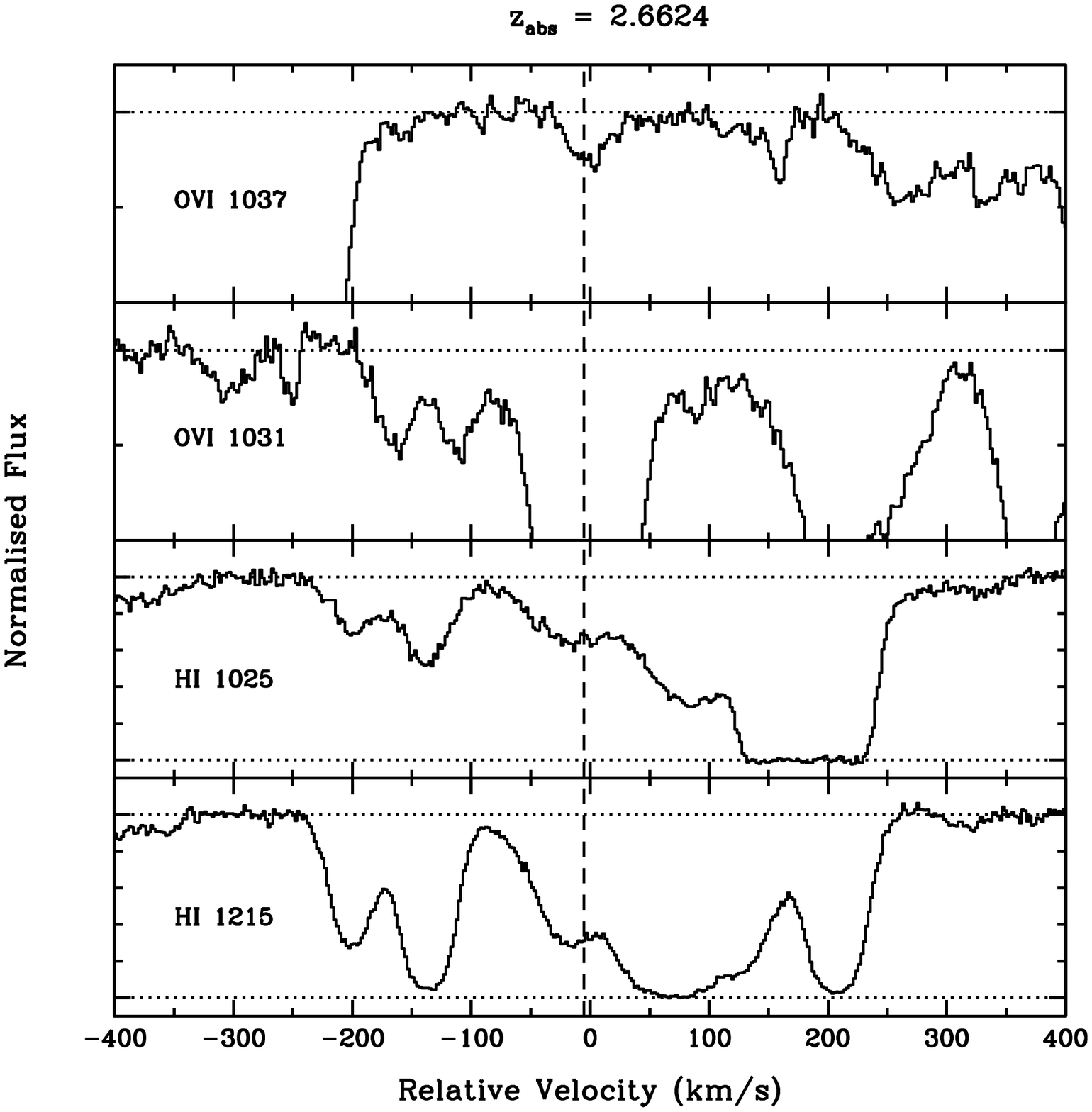}
\includegraphics[height= 8.0cm,width= 8.cm,angle= 0]{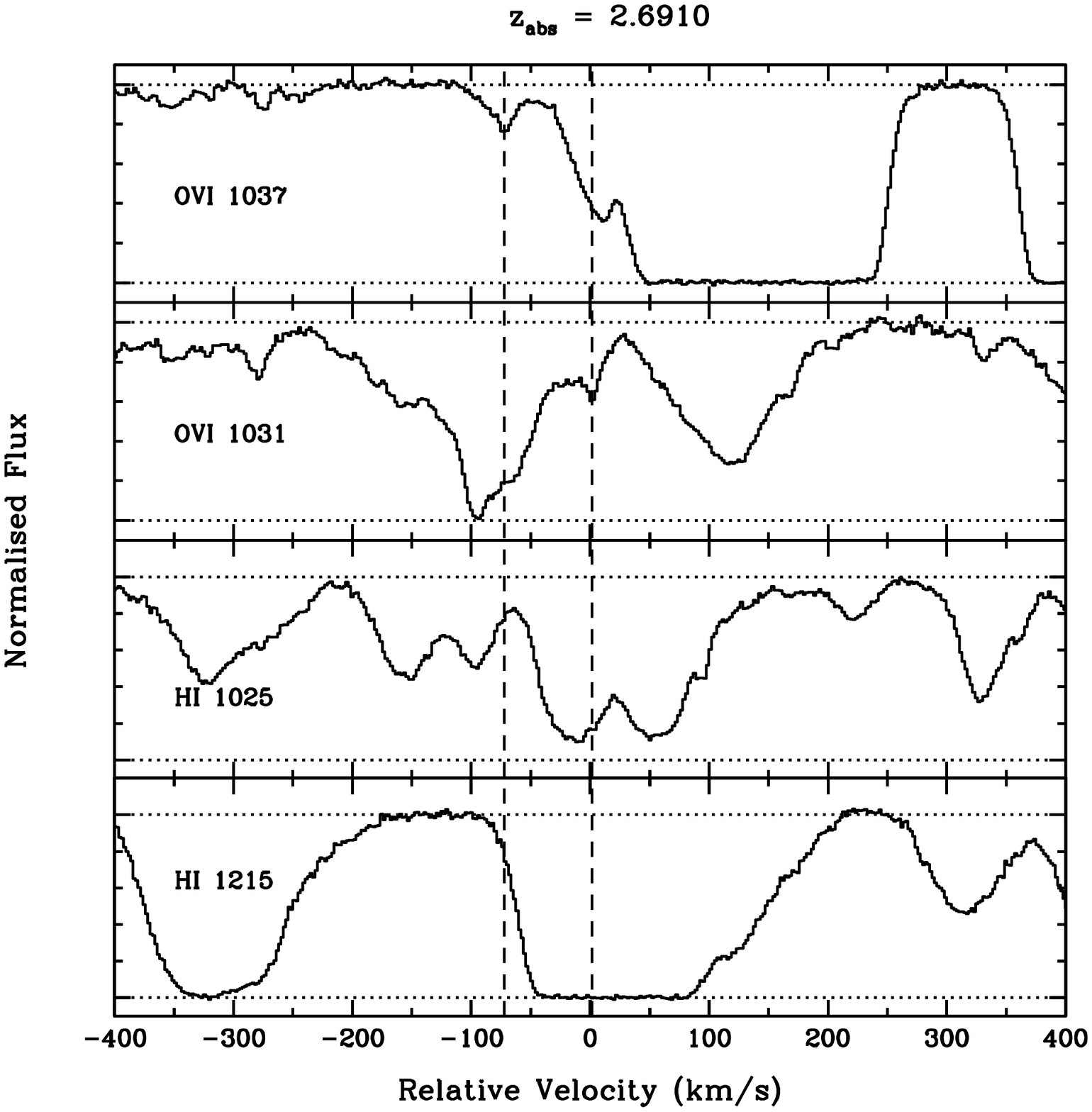}}
}}
\caption{Absorption profiles are shown on a velocity scale for regions with low $\eta$ 
values (see Fig.~1): {\bf D}, {\bf F} (upper-left and right hand-side panels) and  
{\bf H}, {\bf N} (lower left and right hand-side panels). C~{\sc iv} and O~{\sc vi} 
are clearly detected in regions {\bf D} and {\bf F}. In these cases we also overplot the 
best fitted Voigt profile model and indicate individual components with vertical tick marks. 
For regions {\bf H} and {\bf N}, there are only tentative \os\ coincidences. Vertical dashed 
lines in the bottom panels mark the locations of tentative \os\ doublet components. The vertical 
dashed lines in the upper-left panel delineate the region of \os\ absorption. Regions marked 
by `$\times$' are Lyman-$\alpha$ contamination.} 
\label{vplot}
\end{figure*}

Regions {\bf D} and {\bf F} are associated with \cf\ and strong \os\ absorption lines. These are 
the only two \cf\ systems in the redshift range $2.58\le z\le 2.70$ (see top panels in 
Fig.~\ref{vplot}) and we discuss them in detail below.

For region {\bf H}, the wavelength range where possible \os$\lambda$1031 absorption is redshifted 
is strongly blended and only a possible weak line is present at the expected location of 
\os$\lambda$1037 (see bottom-left hand-side panel in Fig.~\ref{vplot}). As no other metal line is 
detected at this redshift we are unable to confirm if this feature is indeed due to \os\ absorption.

For region {\bf N}, while both \os\ regions are strongly blended, the optical depth constraints 
are satisfied at two velocity positions (see Fig.~\ref{vplot}). However, the possible 
O~{\sc vi}$\lambda$1031 feature is also consistent with being C~{\sc iii} absorption at 
\zabs$\sim$2.8972. Similarly, the possible O~{\sc vi}$\lambda$1037 line is blended with 
Lyman-$\beta$ at \zabs~=~2.7306 and O~{\sc vi}$\lambda$1031 at \zabs~=~2.7121. Hence, we cannot 
confirm the presence of \os\ absorption in this region. Note that in region {\bf N} 
(i.e. \zabs = 2.6910) $\eta$ is probably affected by transverse proximity effect from 
QSO J23495-4338 located at redshift $z=2.690$$\pm$0.006, 15~arcmin away from the line of sight of 
interest \citep{Worseck07}.
%
%_______________________ Four OVI systems ___________________________________

Note that we detect \os\ absorption at \zabs = 2.7121, 2.7356 and 2.7456 as well. The \he\ opacity 
is high at \zabs~=~2.7121 and 2.7456 which makes $\eta$\ difficult to estimate. If we scale the 
Voigt-profile fits to the \hi\ absorption to reproduce the \he\ profile, we find $\eta$\ to be 
in the range 10$-$100 and $>$100 for, respectively, the systems at \zabs~=~2.7121 and 2.7456. 
The wavelength range in which the \he\ absorption at \zabs~=~2.7356 is expected to be redshifted 
has been removed by \citet{Zheng04} because of the strong airglow lines so that we cannot estimate 
$\eta$\ for this system. The system at \zabs~=~2.7356 is a known Lyman-limit system. A Voigt-profile 
fit to the \hi\ absorption gives log~$N(\hi)$~=~16.50$\pm$0.28.
%

%_________________________________________________________________________________

\subsection{System at \zabs \ = 2.6498} 

A velocity plot of high ionization metal lines and H~{\sc i} lines from this system is 
shown in the top-right panel of Fig.~\ref{vplot}. Clearly the metal lines are off-centered 
with respect to the \hi\ absorption. In addition, there is a velocity off-set of 2 to 
10~km~s$^{-1}$ between the centroids of the \os\ and \cf\ absorption profiles. Interestingly, 
all the shifts are in the same direction as would be expected in a flow ionized from the same side.
The best fit of the profiles is obtained when we allow for \cf\ component redshifts to be 
independent of that of the \os\ components (see Table~\ref{tab1}).

Doppler parameters are larger for \os\ compared to \cf\ which supports neither pure thermal nor 
pure turbulent broadening. The upper limits on the kinetic temperature of the gas measured from 
the $b$-parameters of \os\ components is 1.4$\times10^{6}$, 8$\times10^4$, 1.2$\times10^5$ 
and 4$\times10^4$~K respectively for components at $-76.2$, $-48.6$, $-20.2$ and $+4.2$ \kms. 
Therefore within the allowed error in $b$-parameters, the \os\ profile allows for the existence 
of high temperature ($T> 10^5$~K) at least in part of the associated gas.

\begin{figure}
\centerline{
\includegraphics[height=9.0cm,width=8.0cm,angle= 0]{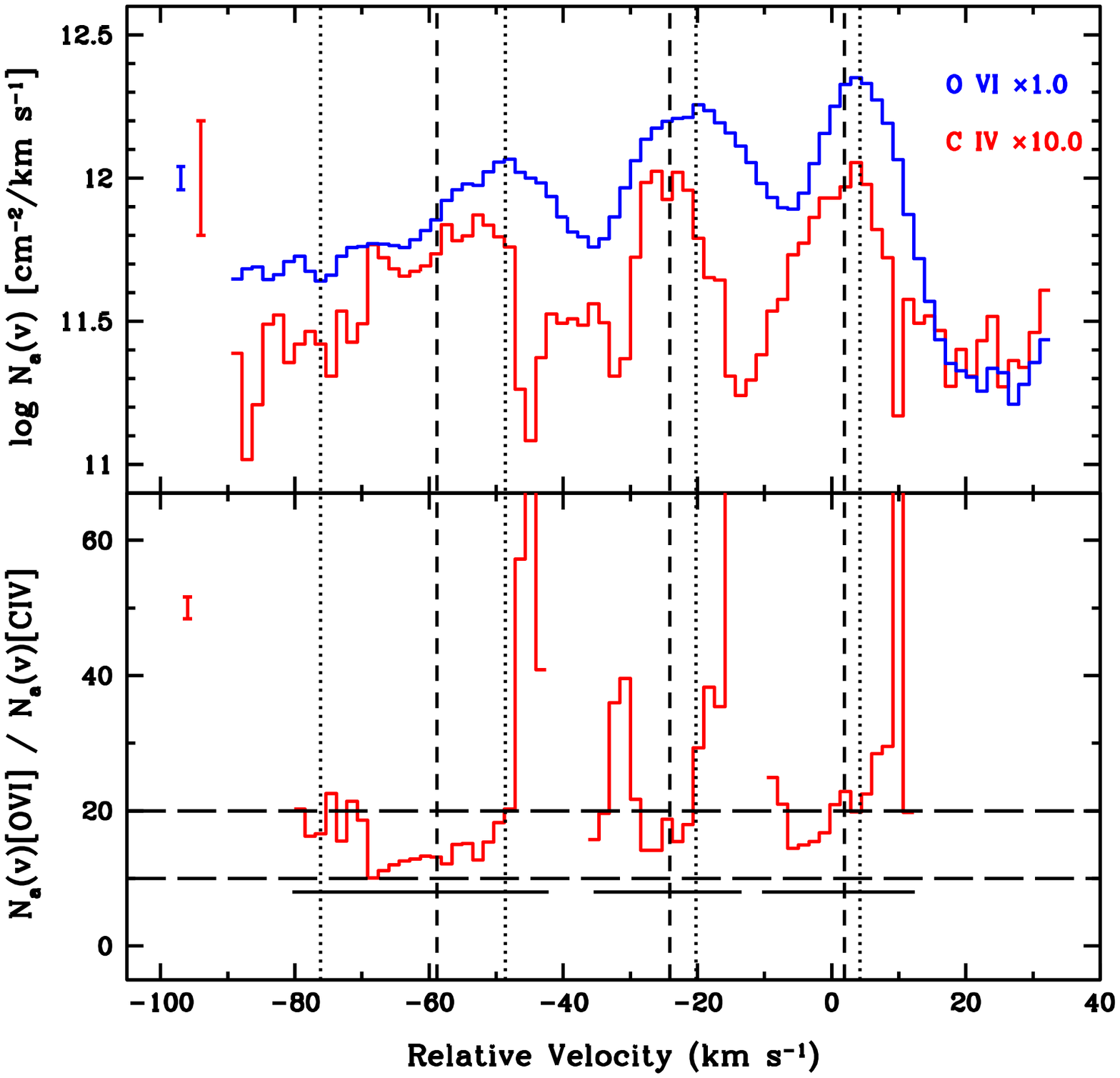}
}
\caption{{\sl Top panel} : Apparent column density profiles of \os\ (in blue) and \cf\ (in red)
              per unit velocity interval versus relative velocity for the system \zabs~ = 2.6498. 
	      For display purpose, \cf\ apparent column densities have been multiplied 
	      by a factor of 10. The mean errors in each pixel for \os\ (in blue errorbar)
	      and \cf\ (in red errorbar) are shown in the left. {\sl Bottom panel} : Apparent column 
	      density ratio through the profile. The errorbar in the left shows corresponding mean 
	      error in each pixel. Vertical dotted lines show the centroids of Voigt components. 
	      The shift between the \os\ and \cf\ centroids is apparent.} 
\label{aod2p64}
\end{figure}

The top panel of Fig.~\ref{aod2p64} shows the apparent column densities of \os\ (in blue) and \cf\ 
(in red) per unit velocity interval versus relative velocity. Since \os\ $\lambda$1037 is heavily 
blended we use only the \os$\lambda$1031 line.
For \cf, we have used the oscillator strength weighted mean of the column densities 
per unit velocity measured from both absorption lines of the doublet. For clarity, we have multiplied 
the \cf\ apparent column density profile by a factor of 10. Vertical dashed and dotted lines show 
the positions of peaks in the optical depth of \cf\ and \os\ respectively. It is apparent that 
the \os\ peaks are shifted compared to the \cf\ ones. 
%
%%___________________ AOD :: z_abs = 2.6498 _____________________________________________

In the lower panel of Fig.~\ref{aod2p64} we plot the ratio of \os\ to \cf\ apparent column 
densities per unit velocity against the relative velocity and find that the ratio varies between 
10 and 20 through the \cf\ absorption profile. The fact that the \os\ absorption profile is 
broader suggests the existence of gas outside the \cf\ profile with \os\ to \cf\ column density 
ratio higher than 20.

The component at $\sim +4.2$~\kms has virtually no detectable H~{\sc i} absorption associated. 
From the Lyman-$\alpha$ line we derive an upper limit of log~$N$(H~{\sc i})~=~12.80 suggesting 
that metallicity is probably high in this component. Indeed, given the low $b$ value of the 
component, it is probable that the gas is  photo-ionized in which case metallicity has to be 
close to solar. For the other three components that coincide with a strong H~{\sc i} absorption 
and it is impossible to quantify the amount of H~{\sc i} absorption associated with them 
individually such that useful metallicity limits can be established. 
%
%%_______________________________________________________________________________________
\begin{table}
\caption{Results of multiple component Voigt profile fitting to the \zabs = 2.6498 system. 
         The parameters of the H~{\sc i} are obtained by keeping the 
         component structure as seen in \os. }
\begin{center}
\begin{tabular}{lclcc}
\hline
 $  v$ (\kms)       &   Ion  &  \hi\ lines used  &   $b$ (\kms)   &   log\,($N$ in \sqcm)   \\  \hline
 $-$214.6 $\pm$ 1.0 &    \hi\  &   \lya            & 16.3 $\pm$ 1.5 &     12.59 $\pm$ 0.03  \\
 $-$150.8 $\pm$ 0.3 &    \hi\  & \lya , \lyb       & 23.6 $\pm$ 0.3 &     14.63 $\pm$ 0.01  \\
  $-$76.2 $\pm$ 0.0 &    \hi\  & \lya , \lyb       & 37.6 $\pm$ 0.9 &     14.91 $\pm$ 0.01  \\
  $-$48.6 $\pm$ 0.0 &    \hi\  & \lya , \lyb       & 23.1 $\pm$ 0.7 &     14.68 $\pm$ 0.02  \\
  $-$20.2 $\pm$ 0.0 &    \hi\  & \lya , \lyb       & 26.4 $\pm$ 1.1 &     13.53 $\pm$ 0.06  \\
\\
  $-$76.2 $\pm$ 3.3 &   \os\   &     & 38.6 $\pm$ 3.9 &     13.56 $\pm$ 0.05  \\
\\
  $-$58.8 $\pm$ 1.0 &   \cf\   &     & 14.5 $\pm$ 1.5 &     12.30 $\pm$ 0.03  \\
  $-$48.6 $\pm$ 0.3 &   \os\   &     &  9.4 $\pm$ 0.9 &     13.16 $\pm$ 0.06  \\
\\
  $-$24.1 $\pm$ 0.6 &   \cf\   &     &  5.9 $\pm$ 1.0 &     12.06 $\pm$ 0.04  \\
  $-$20.2 $\pm$ 0.2 &   \os\   &     & 10.9 $\pm$ 0.4 &     13.55 $\pm$ 0.02  \\
\\
   $+$1.9 $\pm$ 0.4 &   \cf\   &     &  4.8 $\pm$ 0.7 &     12.16 $\pm$ 0.03  \\
   $+$4.2 $\pm$ 0.1 &   \os\   &     &  6.5 $\pm$ 0.2 &     13.49 $\pm$ 0.01  \\
\hline
\end{tabular}
\end{center}
\label{tab1}
\end{table}
%+++++++++++++++++++++++++++++++++++++++++++++++++++++++++++++++++++++++++++++++++

\begin{figure*}
\centerline{
\hbox{
\includegraphics[height= 7.0cm,width= 6.0cm,angle= 0]{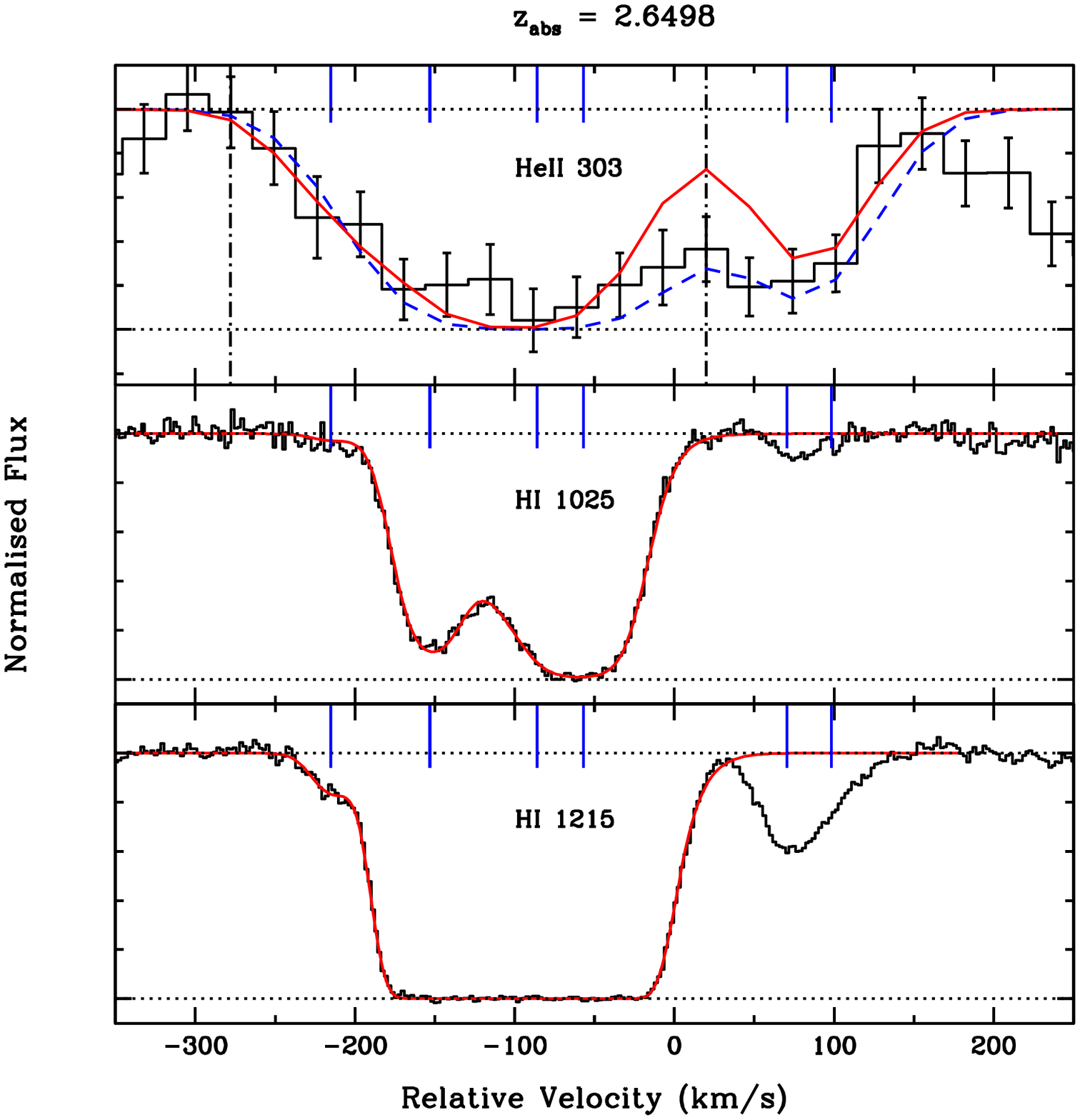}
}
\hbox{
\includegraphics[height= 7.0cm,width= 6.0cm,angle= 0]{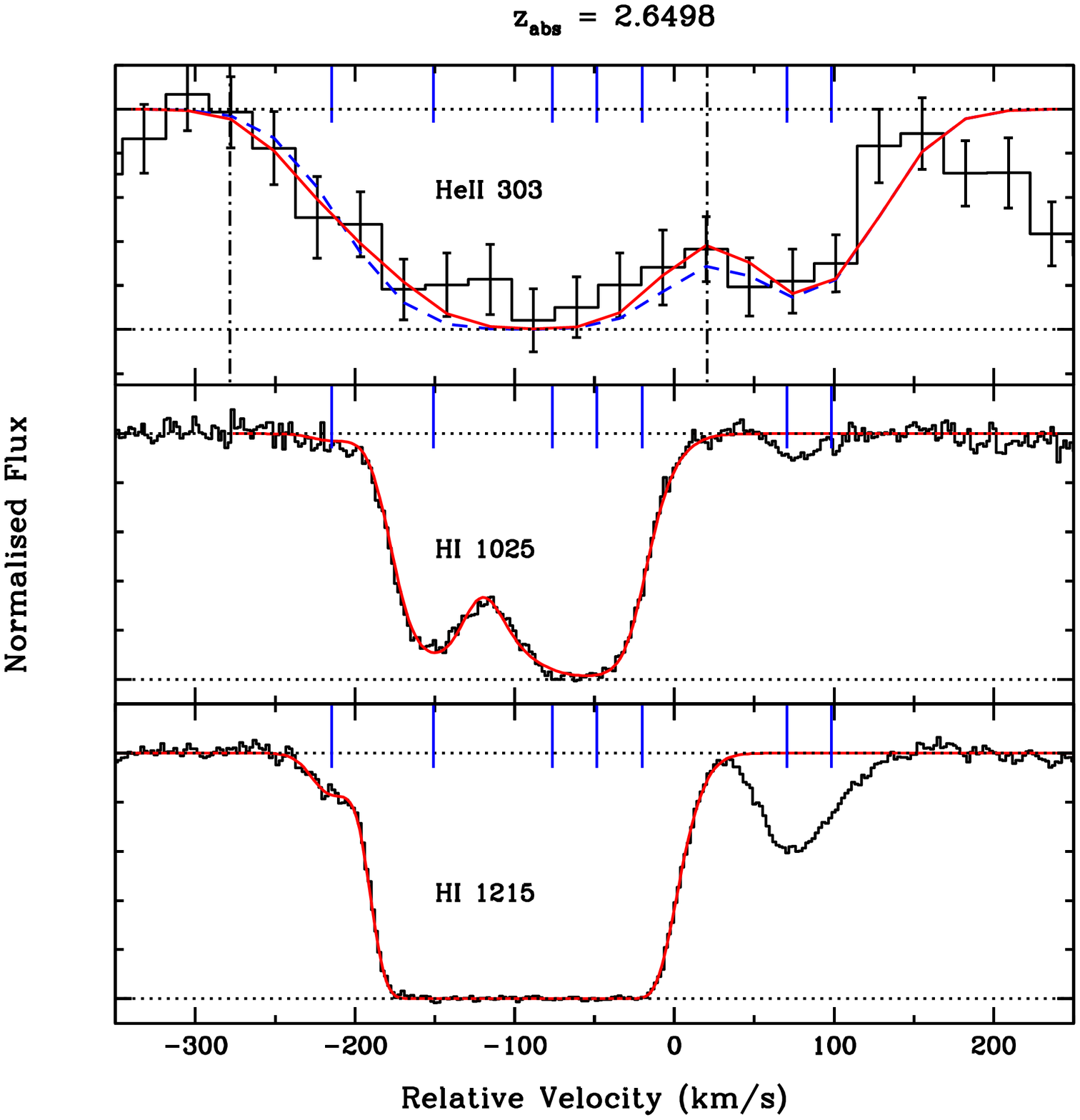}
}
\hbox{
\includegraphics[height= 7.0cm,width= 6.0cm,angle= 0]{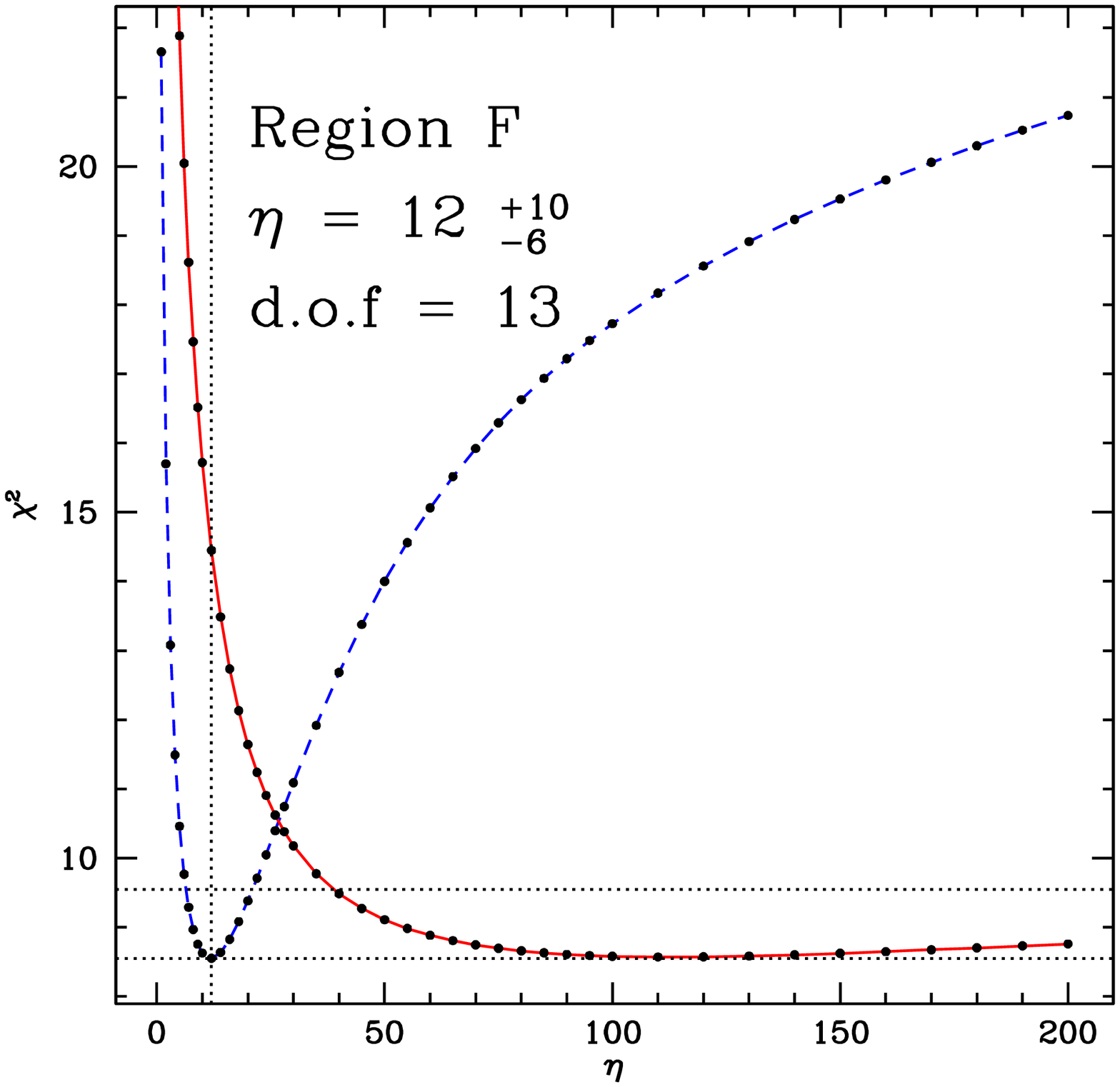}
}
}
\caption{ Fits of H~{\sc i} and He~{\sc ii} absorption in the $z_{\rm abs}$~=~2.6498 system. 
In all panels dashed (blue) and solid (red) curves are for turbulent, 
$\xi=b({\rm HeII})/b({\rm HI})=1$, and thermal, $\xi=0.5$, broadening cases respectively.
{\it Left hand-side column}: Fit using the minimum number (four) of components required to 
fit the H~{\sc i} absorption. The dashed and solid curves in the top panel are the simulated 
\he\ profiles with $\eta \sim$\ 12 and 130 respectively. Only pixels between the two vertical 
dot-dashed lines are used to derive  $\eta$\ by $\chi^{2}$\ minimization. 
{\it Middle column} : Fit with five components imposing H~{\sc i} components at the position of 
the O~{\sc vi} components. The dashed and solid curves in the top panel are the simulated 
\he\ profiles with $\eta \sim$\ 12 and 100 respectively. 
Note that two components around +90~km/s are fitted together to take into account their contribution 
to the He~{\sc ii} absorption. For these two components we take $\eta$~=~65 and pure turbulent 
broadening derived from the analysis of region $F^{\prime}$ (see Fig.~\ref{chisq_plot}). 
{\it Right hand-side column}: $\chi^{2}$\ plot for fits shown in the middle row.
Vertical ticks in middle and left panels indicate the positions of  individual Voigt components.
}
\label{heprof1}
\end{figure*}

We have seen before (Fig.~\ref{chisq_plot}) that the $\chi^{2}$ curve corresponding to the fit
of the He~{\sc ii} absorption shows a marked minimum for $\eta$~=~12 in the case of turbulent 
broadening (i.e ~$\xi$~=~$b_{\he}/ b_{\hi}$ =~1) and no minimum for pure thermal broadening 
($\xi$~=~0.5). In Fig.~\ref{heprof1} we show in the left hand-side row the simulated \he\ profiles 
for $\eta$$\sim$130 and $\xi$~=~0.5 (solid curve) and $\eta \sim$\ 12 and $\xi$~=~1.0 (dashed 
curve). Remember that for these fits we have used the minimum number of Voigt profile 
components without any constraint from the O~{\sc vi} profile. 
It is apparent that the red solid \he\ profile (obtained assuming pure thermal broadening) 
is missing several pixels in the red wing of the region of interest around 0~km/s. This is 
because the $b$ value of the corresponding component (fixed by the H~{\sc i} profile) must 
be much larger to reach this position. If we now add the constraint that H~{\sc i} should be 
associated with the three O~{\sc vi} components, we can reproduce this profile better. 
Indeed, because of the extra component at $v\sim -20$~km/s, pure thermal broadened \he\ 
profile with higher $\eta$\ ($\sim$\ 100) gives an equally good fit (right panel of Fig.~6). 

It seems therefore that if we add a He~{\sc ii} component at the position of the redder O~{\sc vi} 
component, any value of $\eta$ between $\sim$15 and 100 is acceptable. Thus it seems that the 
possible presence of unresolved narrow H~{\sc i} components could be one of causes of low $\eta$ 
measurements. It is a fact however that the main H~{\sc i} components have large $b$ values, 
corresponding to temperatures in excess of 10$^{5}$~K. Therefore it is not impossible that part 
of the gas is at high temperature.

\subsection{System at \zabs \ = 2.6346}
%___________________________________________________________________________________

\begin{table}
\caption{Component Detail of the System at \zabs = 2.6346}
\begin{tabular}{llllc}
\hline
 $  v_0$ (\kms)    &  Ion     & \hi\ lines used   &    $b$ (\kms)   &   log\,($N$ in \sqcm) \\  \hline
$-$140.6 $\pm$ 0.0 &   \hi\   & \lya              &  27.8 $\pm$ 0.6 &     13.41 $\pm$ 0.01  \\

$-$91.6 $\pm$ 0.0  &   \hi\   & \lya ,\lyb ,\lyg  &  22.5 $\pm$ 0.3 &     13.80 $\pm$ 0.01  \\
\\
$-$16.4 $\pm$ 0.2  &   \hi\   & \lya ,\lyb ,\lyg  &  38.2 $\pm$ 0.9 &     14.27 $\pm$ 0.03  \\
                   &   \os\   &                   &  20.3 $\pm$ 8.3 &     12.76 $\pm$ 0.28  \\
                   &   \cf\   &                   &  11.5 $\pm$ 2.4 &     12.55 $\pm$ 0.12  \\
\\
  $-$0.6 $\pm$ 0.7 &   \hi\   & \lya ,\lyb ,\lyg  &  28.3 $\pm$ 2.2 &     14.14 $\pm$ 0.06  \\
                   &   \os\   &                   &  18.9 $\pm$ 0.5 &     14.05 $\pm$ 0.02  \\
                   &   \cf\   &                   &   8.1 $\pm$ 0.6 &     12.83 $\pm$ 0.06  \\
 \\
$+$36.2 $\pm$ 0.7  &   \hi\   & \lya ,\lyb ,\lyg  &  25.7 $\pm$10.4 &     13.24 $\pm$ 0.37  \\
                   &   \os\   &                   &   9.6 $\pm$ 0.8 &     13.37 $\pm$ 0.04  \\
                   &   \cf\   &                   &   7.5 $\pm$ 1.0 &     12.19 $\pm$ 0.04  \\
 \\
$+$48.5 $\pm$ 0.0  &   \hi\   &  \lya ,\lyb ,\lyg &  46.6 $\pm$ 1.5 &     14.32 $\pm$ 0.01  \\
                   &   \os\   &                   &  25.2 $\pm$ 1.6 &     13.61 $\pm$ 0.03  \\
 \\
$+$101.4 $\pm$ 0.0  &   \hi\   &  \lya ,\lyb       &  69.5 $\pm$ 2.0 &     13.69 $\pm$ 0.02  \\
                   &   \os\   &                   &  32.8 $\pm$ 3.5 &     13.35 $\pm$ 0.04  \\
\hline
\end{tabular}
\label{tab2p6346}
\end{table}
%__________________________________________________________________________________

Absorption profiles from this system are shown on a velocity scale in Fig.~\ref{vplot}. Unlike in 
the previous system the velocity range of metal lines falls well within the Lyman-$\alpha$ profile. 
The \os$\lambda$1037 line is  blended with \ Lyman-$\gamma$ at $z$~=~2.8781 and Lyman-$\beta$ \ at
$z$~=~2.6765. Because of this contamination we use the well measured redshifts of \cf\ components
to fit the \os\ doublet. The contributions of the contaminating lines are self-consistently computed 
using other available transitions. In addition to the C~{\sc iv} counterparts, we need two components 
in the red part of the profile to fit the \os \ doublet where there is no \cf\ absorption. H~{\sc i} 
Lyman-$\alpha$,\ Lyman-$\beta$, and Lyman-$\gamma$ lines have been fitted simultaneously imposing 
components at the redshifts of five O~{\sc vi} components. Two extra components are required in the 
blue ($\sim -100$~km/s) to cover the total \hi\ absorption. The details of the fit results are given 
in Table~\ref{tab2p6346}. 

As in the previous system, for the components with both \cf\ and \os\, the \os\ $b$-parameters are 
larger than the \cf\ ones and the column density ratio of \os\ to \cf\ is as high as $\sim 15$. 
The \os\ $b$-parameters correspond to upper limits on the kinetic temperature of 
$4\times 10^5$, $3\times 10^5$ and $9\times 10^4$~K respectively for the components at $-16.4$, $-0.6$ 
and $+36.2$ \kms. In the components where we find only \os\ the ratio of \os\ to \cf\ column densities 
can be higher than 20. These components have broad \os\ lines with $b$-parameters corresponding to 
upper limits of $6\times10^5$ and $10^6$ K respectively for the components at $+48$ and $+101$~\kms. 
The corresponding H~{\sc i} components also have high $b$ values allowing for high temperature 
($\sim 10^5$ K) in the gas associated with these two components. All this suggests a multiphase 
structure in this absorbing gas with the possible existence of a hot phase contributing to most of 
the \os\ absorption. Indeed, the O~{\sc vi} profile is suggestively broad. 
%
%________________________ Helium profile and eta stuff z = 2.6498 _______________________________

%______________________________________________________________________________________

\begin{figure*}{}{}{}{}{}{}
\centerline{
\hbox{
\includegraphics[height= 7.0cm,width= 6.0cm,angle= 0]{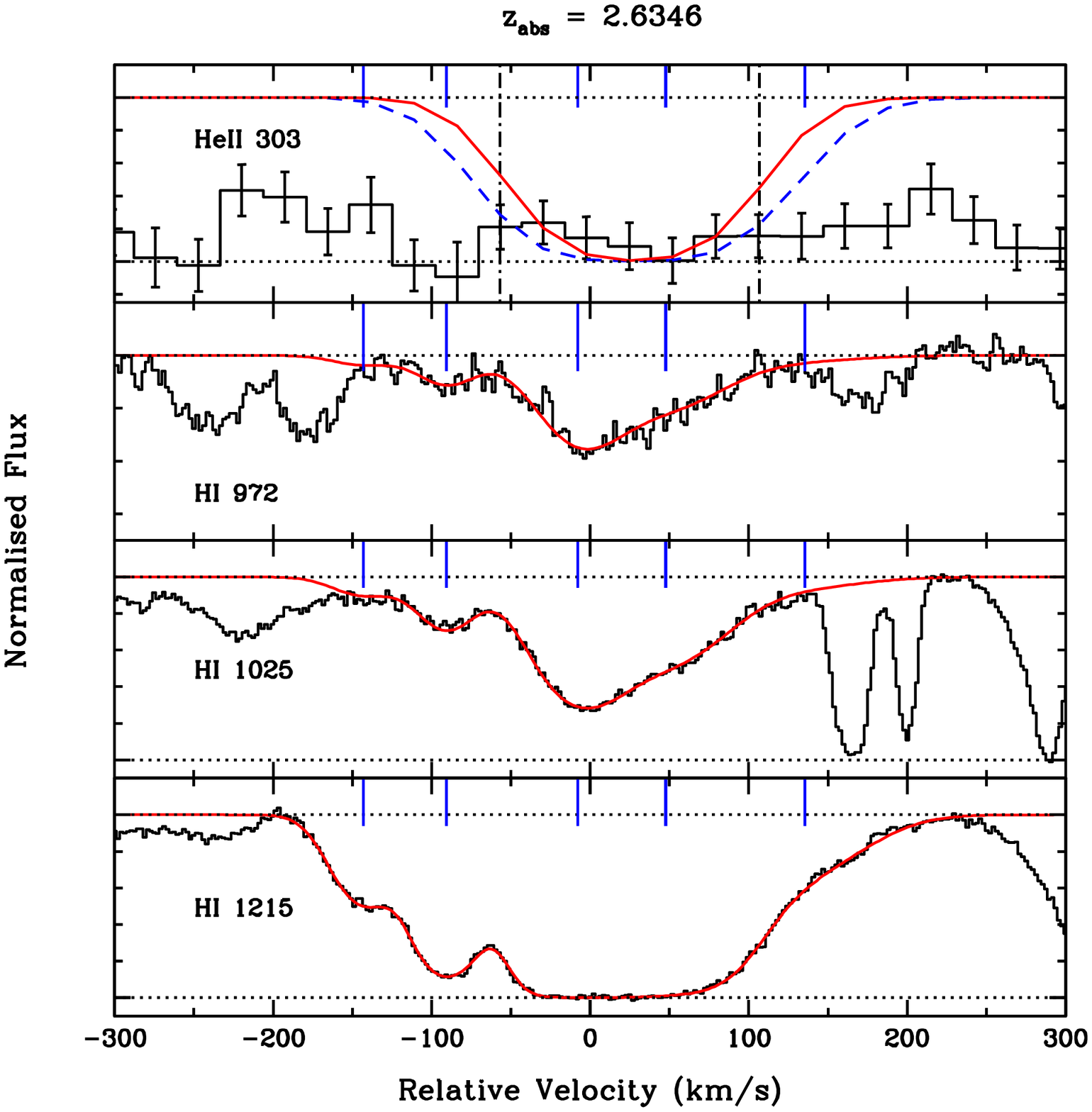}
}
\hbox{
\includegraphics[height= 7.0cm,width= 6.0cm,angle= 0]{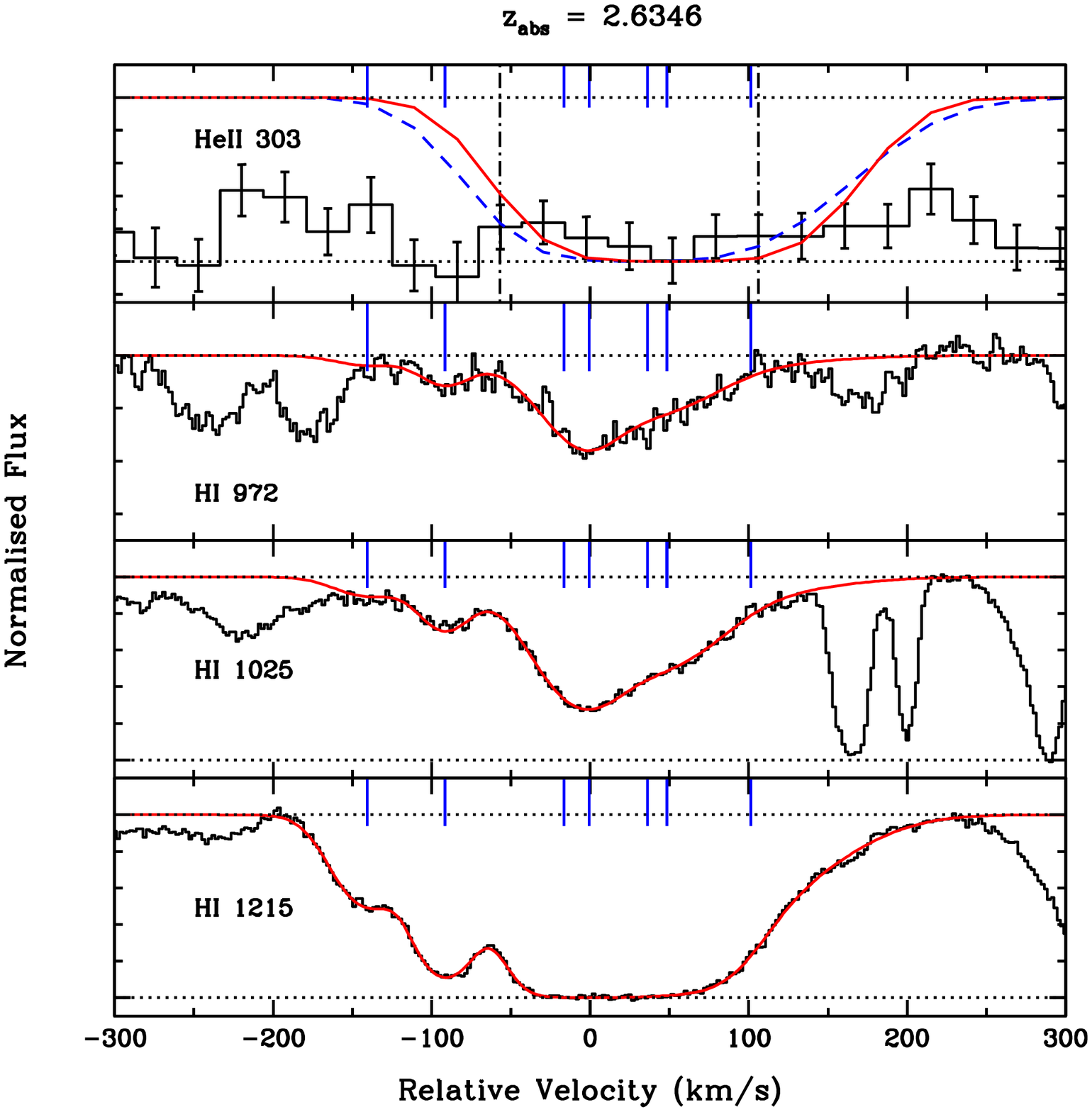}
}
\hbox{
\includegraphics[height= 7.0cm,width= 6.0cm,angle= 0]{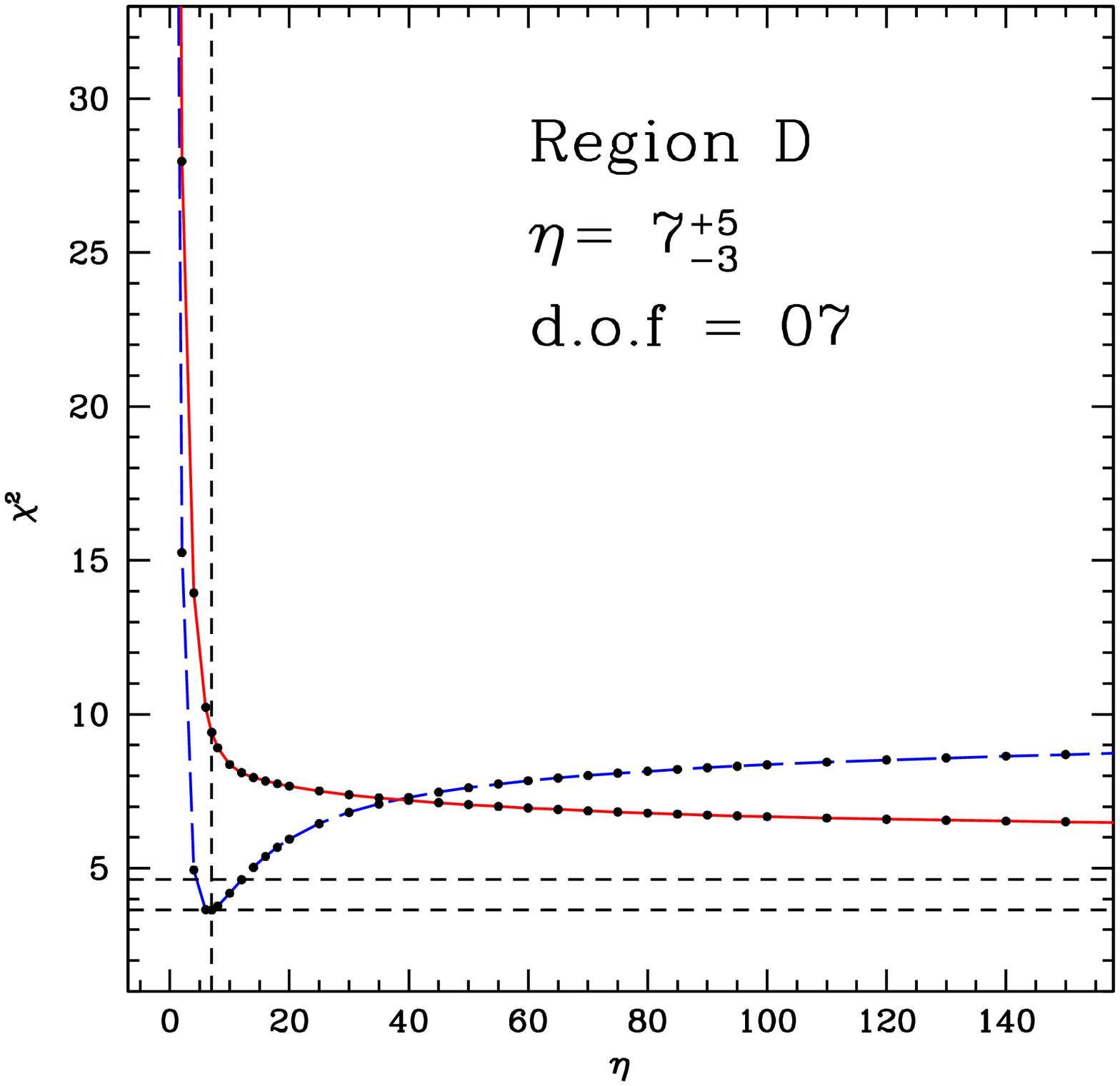}
}
}
\caption{
Fits of H~{\sc i} and He~{\sc ii} absorption in the $z_{\rm abs}$~=~2.6346 system. In all panels 
dashed (blue) and solid (red) curves are for turbulent, $\xi=b({\rm HeII})/b({\rm HI})=1$, and 
thermal, $\xi=0.5$, broadening cases respectively.
{\it Left hand-side column}: Fit using the minimum number (five) of components required to fit 
the H~{\sc i} absorption. The dashed and solid curves in the top panel are the simulated \he\ 
profiles with $\eta \sim$\ 12 and 160 respectively. Only pixels between the two vertical dot-dashed 
lines are used to derive  $\eta$\ by $\chi^{2}$\ minimization. 
{\it Middle column}: Fit with seven components imposing H~{\sc i} components at the position of 
the O~{\sc vi} components. The dashed and solid curves in the top panel are the simulated \he\ 
profiles with $\eta \sim$\ 7 and 140 respectively.
{\it Right hand-side column}: $\chi^{2}$\ plot for fits shown in the middle row.
Vertical ticks in middle and left panels indicate the positions of individual Voigt components.
}
\label{heprof2}
\end{figure*}

\begin{figure}
\centerline{
\hbox{
\includegraphics[height= 9.0cm,width= 8.0cm,angle= 0]{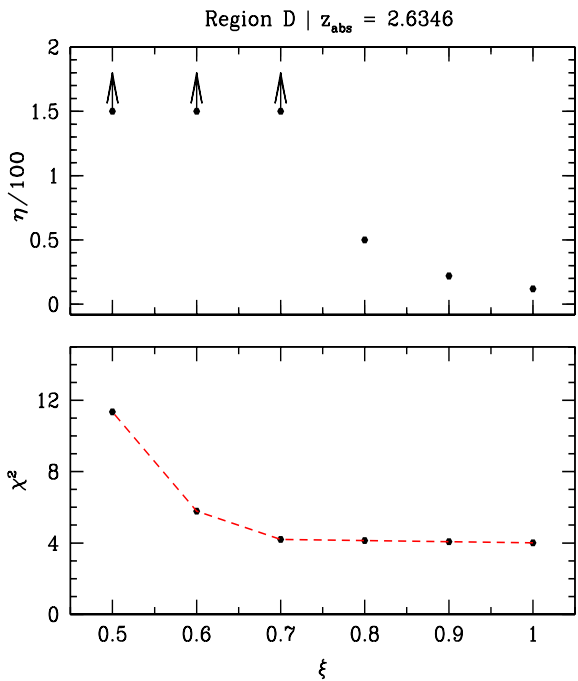}
}
}
\caption{System at \zabs = 2.6346. {\it Bottom:} Minimum $\chi^2$ as a function of $\xi$ 
in the case of H~{\sc i} fitted with minimum number of components. 
{\it Top:} Best fitted value of $\eta$ for different values of $\xi$.} 
\label{xichisq1}
\end{figure}

We fitted the H~{\sc i} and He~{\sc ii} profiles in the two extreme cases of pure turbulent 
and pure thermal broadening, considering both components from the fit of the H~{\sc i} profile 
only and from the fit of metal lines. Results are given in Fig~\ref{heprof2} and Table~2. 
We notice from right panel of Fig~\ref{heprof2} that even when we tie the H~{\sc i} components 
to \os\ components the best fitted $\chi^2$ is obtained for the pure turbulent case with 
low $\eta$. However, reality probably corresponds to an intermediate case with $\xi$ between 0.5 and 1. 
In the bottom panel of Fig.~\ref{xichisq1} we plot the minimum $\chi^2$ value obtained for different 
values of $\xi$. Even though the best fitted $\chi^2$ value is obtained for $\xi$~=~1 the curve is flat 
and the 1$\sigma$ range is $\xi\ge0.6$. As can be seen in the top panel of the figure, this can 
accommodate a wide range of $\eta$. Therefore, in this system also high $\eta$ values are acceptable 
although H~{\sc i} and O~{\sc vi} absorption profiles are broad and highly suggestive of a gas 
with temperature higher than the typical photo-ionization temperature (i.e few $10^4$ K).
%
%__________________________ Chi2 stuff for 2.6346 _____________________________________

%================================= SECTION :: MODELS =====================================================

\section{Models}

Given the particularities of the systems singled out by the presence of O~{\sc vi} absorption, 
possible low $\eta$ values and high O~{\sc vi}/C~{\sc iv} ratios, we have constructed models to 
test the different mechanisms that could induce such properties. It is well known that 
photo-ionization by a power-law spectrum with appropriate slope can yield low $\eta$ values. 
This would require the presence of local sources of hard photons \citep[see][]{Shull04}. 
Observations by \citet{Worseck07} seem to show however that there is no QSO present in the 
vicinity of the two absorbers considered in the previous Section. While this observation does not 
rule out a QSO emission highly beamed perpendicular to the line of sight or a short lived QSO emission 
in the vicinity of the absorbers, we explore alternate explanations for low $\eta$ in the \os\ 
absorbers. Therefore, in the following we present the results of models of a hot gas embedded in the 
meta-galactic UV background.

\begin{figure}
\centerline{
\vbox{\includegraphics[height=8.0cm,width=8.0cm,angle= 0]{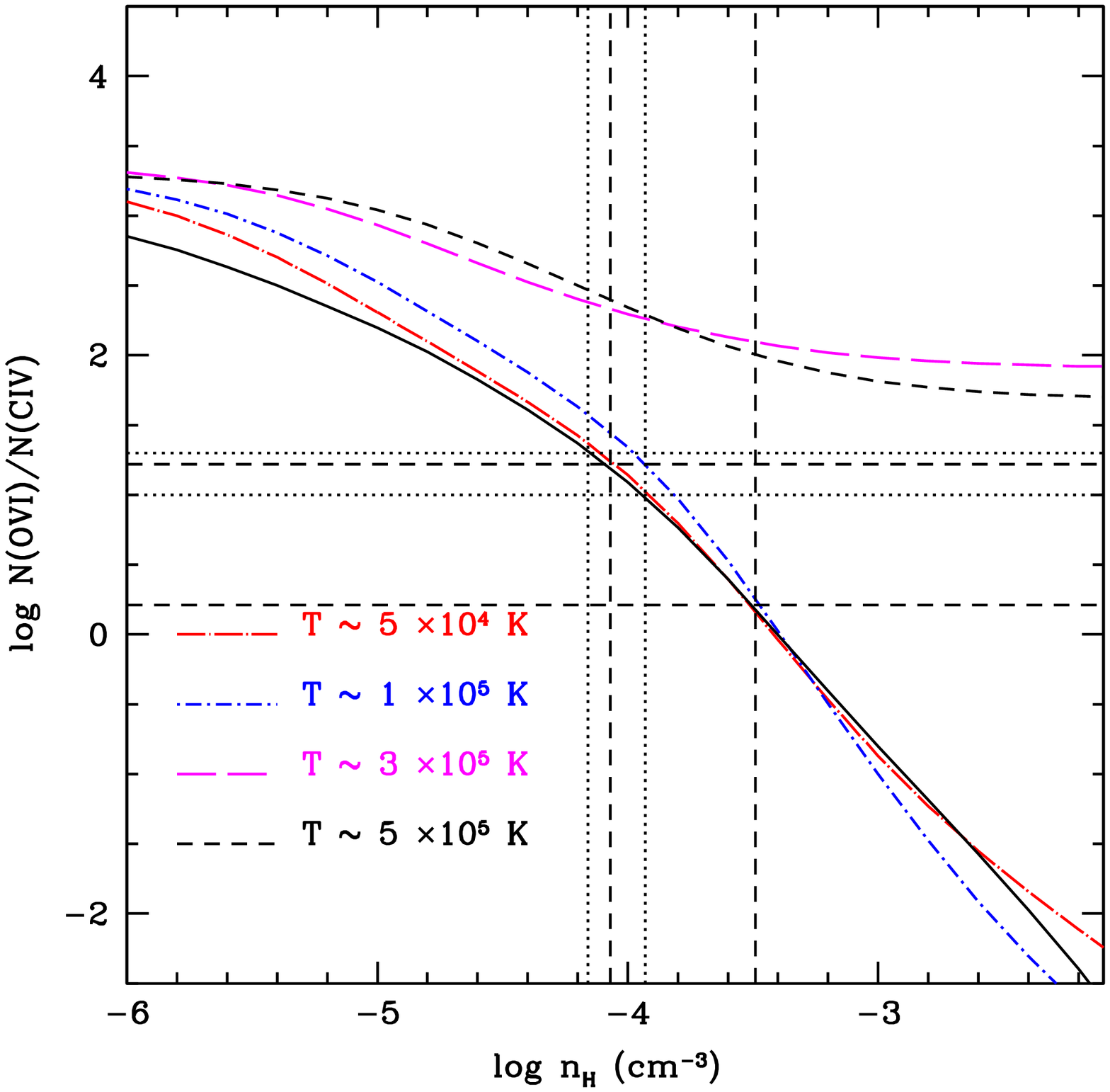}
\includegraphics[height=8.0cm,width=8.0cm,angle= 0]{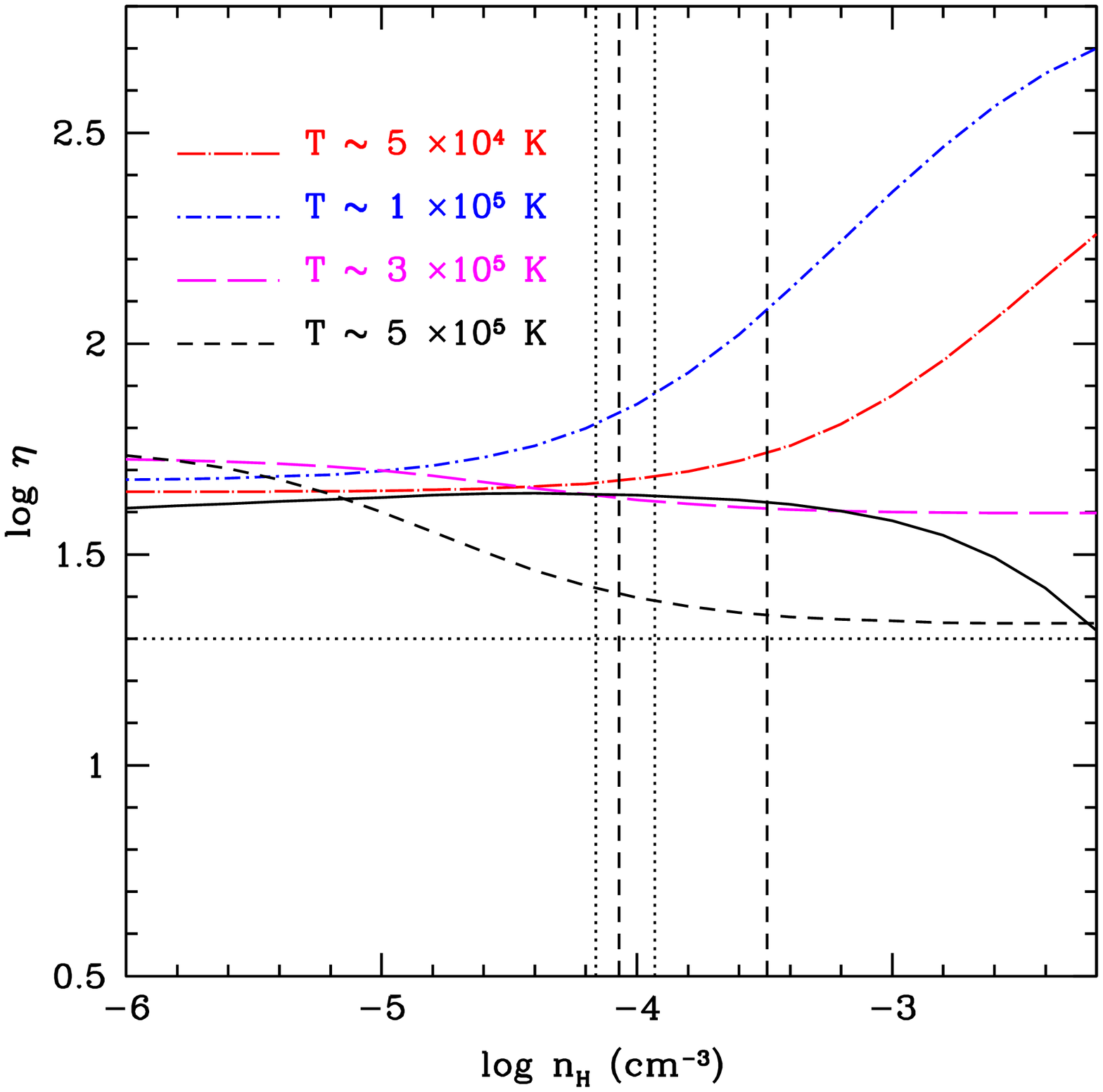}
}}
\caption{{\sl Top panel}: The \os\ to \cf\ column density ratio versus hydrogen density. 
              The different curves correspond to the results of optically thin CLOUDY models with 
	      constant temperature except the solid black curve which is for thermal equilibrium. 
              The gas is assumed of metallicity [C/H]~=~$-1.0$ \& solar relative abundances and 
	      exposed to QSO dominated HM05 ionizing flux. The horizontal dotted and dashed lines 
	      show the range of observed column density ratios for the system at \zabs = 2.6498 
	      and 2.6346 respectively. The vertical dotted and dashed lines indicate the 
	      corresponding allowed range in hydrogen density for photo-ionization equilibrium.
	 }
\label{constT_model}
\end{figure}

We use the photo-ionization code Cloudy (v07.02; \citet{Ferland98}) to derive the ionization structure 
in a gas with fixed temperature (therefore {\sl not} controlled by photo-ionization).
This will make it possible to discuss at the same time both extreme situations (collisional ionization 
and photo-ionization) but also the intermediate situation of high-temperature gas with a contribution 
of photo-ionization. For comparison, we also show results from the model in which the temperature is 
the consequence of thermal equilibrium under photo-ionization. The calculations are made in the 
optically thin case. We use the Haardt and Madau (2005) background spectrum dominated by QSOs. We 
assume relative solar abundances and  [C/H]~=~$-1.0$. In the top panel of Fig.~\ref{constT_model}, we 
plot the variation of the \os\ to \cf\ ratio with hydrogen density. The solid black line is the result 
of model calculations where temperature is calculated by CLOUDY assuming photo-ionization equilibrium. 
Other lines are for temperatures in the range 5$\times$10$^{4}$$-$5$\times$10$^{5}$~K. It is to be 
remembered that when pure collisional excitation is considered the fraction of He~{\sc ii} is maximum 
when 4.5~$\le log~{\rm T(K)} \le$~4.9 and in the case of \os\ it is $T\sim$~3$\times$$10^5$~K 
\citep{Gnat07}. 
At low temperature (say $T\le5\times 10^{4}$~K) the ionization is dominated by photo-ionization. As 
expected the transition between photo-ionization dominated and collisional ionization dominated 
regimes happens at $T\sim 10^{5}$~K. 

The horizontal dotted lines show the range of observed \os\ to \cf\ column density ratios (between 
10 and 20) seen in the \cf\ components of the two systems discussed above. This range is well 
reproduced by models with T$\le 10^5$ K for a typical density of 10$^{-4}$~cm$^{-3}$. However the 
higher \os\ to \cf\ ratio inferred in the velocity range (or Voigt profile components) where only 
\os\ is detected needs either low density (and low temperature) photo-ionized gas or high density 
(i.e $\ge 10^{-3}$~cm$^{-3}$) hot gas (T$> 10^5$ K) where collisions begin to play a role. 
Interestingly such high temperatures are not ruled out by the $b$-parameters of \os\ components 
(see discussions in the previous Section).
%____________________________ XI VS   ETA __________________________________________________
%

%________________________ Constant Temp. Photo. Model __________________________________
%________________________________________________________________________________________

In the bottom panel of Fig.~\ref{constT_model} we plot $\eta$ as predicted by the models versus
the hydrogen density. It is apparent that low $\eta$ values (i.e $\le 60$) are only possible 
for $T>10^{5}$~K. Available data on H~{\sc i} and He~{\sc ii} profiles allow for the existence 
of such hot gas that would also produce the component with high \os\ to \cf\ column density ratio 
(i.e $N$(\os)/$N$(\cf)$\ge$20). It is apparent from Fig.~4 that the absorption profiles 
indicate higher Doppler parameters going from C~{\sc iv} to O~{\sc vi} to H~{\sc i}. 
This has already been noted for C~{\sc iv} and O~{\sc vi} by Fox et al. (2007) and interpreted 
as the existence of a hot phase. We note that the $b$ values measured for the strongest H~{\sc i} 
components in the two systems (38.6~km/s at $z_{\rm abs}$~=~2.6498 and 46.6 and 69.5~km~s$^{-1}$ at 
$z_{\rm abs}$~=~2.6346, see Tables~1 and 2) are consistent with a temperature, $T\ge10^{5}$~K and  
it is apparent from the absorption profiles that larger $b$ values could be accommodated.

If the low $\eta$ values were to be confirmed, we would favor a scenario where the absorbing gas 
is a multiphase medium in which photo-ionized gas components coexist with a wide range of density 
and temperature. While most of the metal absorption traced by \cf\ comes from relatively cold 
(i.e T$\le 10^5$ K) gas part of \os\ and predominant contributions of  H~{\sc i} and He~{\sc ii} could 
be due to a hot phase ($T>10^{5}$~K). There is evidence for the existence of multiphase media 
in the low-z \os\ absorbers \citep{Tripp08} and \os\ absorption associated with high-$z$ 
DLAs \citep{Fox07b}.
%

%================================= SEction :: Conclusion =================================================

\section{Conclusions}
We have reanalyzed the line of sight towards the high redshift ($z_{\rm em}$~=~2.885) bright quasar 
QSO HE~2347$-$4342 and measured the parameter $\eta$~=~$N$(He~{\sc ii})/$N$(H~{\sc i}) in the 
Lyman-$\alpha$ forest using Voigt-profile fitting of the H~{\sc i} transitions in the Lyman series. 
As in previous studies, we find that $\eta > 50$ in most of the Lyman-$\alpha$ forest except in four 
regions where it is much smaller ($\eta \sim 10-20$).

We detect O~{\sc vi} absorption associated with two of these regions 
(at $z_{\rm abs}$~=~2.6346 and 2.6498). The corresponding wavelength ranges for the two other regions 
are too blended to reach any firm conclusion on the presence of associated O~{\sc vi} absorption.
We observe that the $z_{\rm abs}$~=~2.6346 system is a usual system with the metals located at the 
center of the H~{\sc i} profile whereas the $z_{\rm abs}$~=~2.6498 system has the metals displaced in 
the red wing of the H~{\sc i} absorption but moreover, with the C~{\sc iv} profile systematically 
shifted compared to O~{\sc vi}. Doppler parameters of the well-defined C~{\sc iv} components rule out 
the fact that the associated gas is hot and favor the idea that this gas is photo-ionized. We show that 
if we constrain the fit of the H~{\sc i} and/or He~{\sc ii} absorption profiles with the presence of 
metal components, we can accommodate $\eta$ values in the range 15--100 in these systems assuming 
broadening is intermediate between pure thermal and pure collisional.

We construct constant density photo-ionized models and show that while simple photo-ionization 
models reproduce the  observed $N$(O~{\sc vi})/$N$(C~{\sc iv}) ratio for a range of density, they 
fail to produce low $\eta$ values. On the contrary, models with high temperature 
(i.e T $\ge 10^5$~K) can produce low values of $\eta$. 
The Doppler parameters of the strongest H~{\sc i} components are consistent with such a temperature. 
In addition, the higher $b$ values observed for \os\ compared to C~{\sc iv} and the existence of 
\os\ alone components suggest a multiphase nature of the absorbing region. Therefore, if low 
$\eta$ values were to be confirmed, we would favor a multi-phase model in which most of the gas in 
the regions of low $\eta$ is at high temperature ($>$10$^5$~K) but the metals and in particular 
C~{\sc iv} are located in lower temperature photo-ionized and probably transient regions. As the high 
temperature gas can not be  produced by photo-ionization, we expect the \os\ systems with low $\eta$ 
to be associated with galaxies. Therefore, deep search for Lyman break galaxies at these redshifts 
may be interesting to perform in these fields.

\section*{Acknowledgment}
We wish to thank Dr. Zheng for providing the FUSE data and the referee Dr. Williger for 
useful comments. SM thanks CSIR for providing support for this work.
RS thanks University Paris 6 and IAP for an invitation as Professeur Associ\'e.

%==========================================
%----------------- Bibliography and bibfile
\def\aj{AJ}%
\def\actaa{Acta Astron.}%
\def\araa{ARA\&A}%
\def\apj{ApJ}%
\def\apjl{ApJ}%
\def\apjs{ApJS}%
\def\ao{Appl.~Opt.}%
\def\apss{Ap\&SS}%
\def\aap{A\&A}%
\def\aapr{A\&A~Rev.}%
\def\aaps{A\&AS}%
\def\azh{AZh}%
\def\baas{BAAS}%
\def\bac{Bull. astr. Inst. Czechosl.}%
\def\caa{Chinese Astron. Astrophys.}%
\def\cjaa{Chinese J. Astron. Astrophys.}%
\def\icarus{Icarus}%
\def\jcap{J. Cosmology Astropart. Phys.}%
\def\jrasc{JRASC}%
\def\mnras{MNRAS}%
\def\memras{MmRAS}%
\def\na{New A}%
\def\nar{New A Rev.}%
\def\pasa{PASA}%
\def\pra{Phys.~Rev.~A}%
\def\prb{Phys.~Rev.~B}%
\def\prc{Phys.~Rev.~C}%
\def\prd{Phys.~Rev.~D}%
\def\pre{Phys.~Rev.~E}%
\def\prl{Phys.~Rev.~Lett.}%
\def\pasp{PASP}%
\def\pasj{PASJ}%
\def\qjras{QJRAS}%
\def\rmxaa{Rev. Mexicana Astron. Astrofis.}%
\def\skytel{S\&T}%
\def\solphys{Sol.~Phys.}%
\def\sovast{Soviet~Ast.}%
\def\ssr{Space~Sci.~Rev.}%
\def\zap{ZAp}%
\def\nat{Nature}%
\def\iaucirc{IAU~Circ.}%
\def\aplett{Astrophys.~Lett.}%
\def\apspr{Astrophys.~Space~Phys.~Res.}%
\def\bain{Bull.~Astron.~Inst.~Netherlands}%
\def\fcp{Fund.~Cosmic~Phys.}%
\def\gca{Geochim.~Cosmochim.~Acta}%
\def\grl{Geophys.~Res.~Lett.}%
\def\jcp{J.~Chem.~Phys.}%
\def\jgr{J.~Geophys.~Res.}%
\def\jqsrt{J.~Quant.~Spec.~Radiat.~Transf.}%
\def\memsai{Mem.~Soc.~Astron.~Italiana}%
\def\nphysa{Nucl.~Phys.~A}%
\def\physrep{Phys.~Rep.}%
\def\physscr{Phys.~Scr}%
\def\planss{Planet.~Space~Sci.}%
\def\procspie{Proc.~SPIE}%
\let\astap=\aap
\let\apjlett=\apjl
\let\apjsupp=\apjs
\let\applopt=\ao
\bibliographystyle{mn}
\bibliography{mybib}

\begin{thebibliography}{37}
\expandafter\ifx\csname natexlab\endcsname\relax\def\natexlab#1{#1}\fi

\bibitem[{{Aguirre} {et~al.}(2008){Aguirre}, {Dow-Hygelund}, {Schaye}, \&
  {Theuns}}]{Aguirre08}
{Aguirre}, A., {Dow-Hygelund}, C., {Schaye}, J., \& {Theuns}, T., 2008, \apj,
  689, 851

\bibitem[{{Aracil} {et~al.}(2004){Aracil}, {Petitjean}, {Pichon}, \&
  {Bergeron}}]{Aracil04}
{Aracil}, B., {Petitjean}, P., {Pichon}, C., \& {Bergeron}, J., 2004, \aap,
  419, 811

\bibitem[{{Ballester} {et~al.}(2000){Ballester}, {Modigliani}, {Boitquin},
  {Cristiani}, {Hanuschik}, {Kauffer}, \& {Wolf}}]{Ballester00}
{Ballester}, P., {Modigliani}, A., {Boitquin}, O., {Cristiani}, S.,
  {Hanuschik}, R., {Kauffer}, A., \& {Wolf}, S., 2000, The Messenger, 101, 31

\bibitem[{{Bergeron} {et~al.}(2002){Bergeron}, {Aracil}, {Petitjean}, \&
  {Pichon}}]{Bergeron02}
{Bergeron}, J., {Aracil}, B., {Petitjean}, P., \& {Pichon}, C., 2002, \aap,
  396, L11

\bibitem[{{Bergeron} {et~al.}(2004){Bergeron}, {Petitjean}, {Aracil}, {Pichon},
  {Scannapieco}, {Srianand}, {Boisse}, {Carswell}, {Chand}, {Cristiani},
  {Ferrara}, {Haehnelt}, {Hughes}, {Kim}, {Ledoux}, {Richter}, \&
  {Viel}}]{Bergeron04}
{Bergeron}, J., {Petitjean}, P., {Aracil}, B., {et~al.}, 2004, The Messenger,
  118, 40

\bibitem[{{Bouch{\'e}} {et~al.}(2007){Bouch{\'e}}, {Lehnert}, {Aguirre},
  {P{\'e}roux}, \& {Bergeron}}]{Bouche07}
{Bouch{\'e}}, N., {Lehnert}, M.~D., {Aguirre}, A., {P{\'e}roux}, C., \&
  {Bergeron}, J., 2007, \mnras, 378, 525

\bibitem[{{Bouch{\'e}} {et~al.}(2006){Bouch{\'e}}, {Lehnert}, \&
  {P{\'e}roux}}]{Bouche06}
{Bouch{\'e}}, N., {Lehnert}, M.~D., \& {P{\'e}roux}, C., 2006, \mnras, 367, L16

\bibitem[{{Chand} {et~al.}(2004){Chand}, {Srianand}, {Petitjean}, \&
  {Aracil}}]{Chand04}
{Chand}, H., {Srianand}, R., {Petitjean}, P., \& {Aracil}, B., 2004, \aap, 417,
  853

\bibitem[{{Dav{\'e}} {et~al.}(2001){Dav{\'e}}, {Cen}, {Ostriker}, {Bryan},
  {Hernquist}, {Katz}, {Weinberg}, {Norman}, \& {O'Shea}}]{Dave01}
{Dav{\'e}}, R., {Cen}, R., {Ostriker}, J.~P., {et~al.}, 2001, \apj, 552, 473

\bibitem[{{Dekker} {et~al.}(2000){Dekker}, {D'Odorico}, {Kaufer}, {Delabre}, \&
  {Kotzlowski}}]{Dekker00}
{Dekker}, H., {D'Odorico}, S., {Kaufer}, A., {Delabre}, B., \& {Kotzlowski},
  H., 2000, in Proc. SPIE Vol. 4008, p. 534-545, Optical and IR Telescope
  Instrumentation and Detectors, Masanori Iye; Alan F. Moorwood; Eds., pp.
  534--545

\bibitem[{{Edl{\'e}n}(1966)}]{Edlen66}
{Edl{\'e}n}, B., 1966, Metrologia, 2, 71

\bibitem[{{Ellison} {et~al.}(2000){Ellison}, {Songaila}, {Schaye}, \&
  {Pettini}}]{Ellison00}
{Ellison}, S.~L., {Songaila}, A., {Schaye}, J., \& {Pettini}, M., 2000, \aj,
  120, 1175

\bibitem[{{Fang} \& {Bryan}(2001)}]{Fang01}
{Fang}, T. \& {Bryan}, G.~L., 2001, \apjl, 561, L31

\bibitem[{{Fardal} {et~al.}(1998){Fardal}, {Giroux}, \& {Shull}}]{Fardal98}
{Fardal}, M.~A., {Giroux}, M.~L., \& {Shull}, J.~M., 1998, \aj, 115, 2206

\bibitem[{{Fechner} \& {Reimers}(2007)}]{Fechner07a}
{Fechner}, C. \& {Reimers}, D., 2007, \aap, 461, 847

\bibitem[{{Ferland} {et~al.}(1998){Ferland}, {Korista}, {Verner}, {Ferguson},
  {Kingdon}, \& {Verner}}]{Ferland98}
{Ferland}, G.~J., {Korista}, K.~T., {Verner}, D.~A., {Ferguson}, J.~W.,
  {Kingdon}, J.~B., \& {Verner}, E.~M., 1998, \pasp, 110, 761

\bibitem[{{Fox} {et~al.}(2007){Fox}, {Petitjean}, {Ledoux}, \&
  {Srianand}}]{Fox07b}
{Fox}, A.~J., {Petitjean}, P., {Ledoux}, C., \& {Srianand}, R., 2007, \aap,
  465, 171

\bibitem[{{Gnat} \& {Sternberg}(2007)}]{Gnat07}
{Gnat}, O. \& {Sternberg}, A., 2007, \apjs, 168, 213

\bibitem[{{Haardt} \& {Madau}(1996)}]{Haardt96}
{Haardt}, F. \& {Madau}, P., 1996, \apj, 461, 20

\bibitem[{{Kang} {et~al.}(2005){Kang}, {Ryu}, {Cen}, \& {Song}}]{Kang05}
{Kang}, H., {Ryu}, D., {Cen}, R., \& {Song}, D., 2005, \apj, 620, 21

\bibitem[{{Khare} {et~al.}(1997){Khare}, {Srianand}, {York}, {Green}, {Welty},
  {Huang}, \& {Bechtold}}]{Khare97}
{Khare}, P., {Srianand}, R., {York}, D.~G., {Green}, R., {Welty}, D., {Huang},
  K., \& {Bechtold}, J., 1997, \mnras, 285, 167

\bibitem[{{Kriss} {et~al.}(2001){Kriss}, {Shull}, {Oegerle}, {Zheng},
  {Davidsen}, {Songaila}, {Tumlinson}, {Cowie}, {Deharveng}, {Friedman},
  {Giroux}, {Green}, {Hutchings}, {Jenkins}, {Kruk}, {Moos}, {Morton},
  {Sembach}, \& {Tripp}}]{Kriss01}
{Kriss}, G.~A., {Shull}, J.~M., {Oegerle}, W., {et~al.}, 2001, Science, 293,
  1112

\bibitem[{{Oppenheimer} \& {Dav{\'e}}(2009)}]{Oppenheimer09}
{Oppenheimer}, B.~D. \& {Dav{\'e}}, R., 2009, \mnras, 395, 1875

\bibitem[{{Pieri} {et~al.}(2010){Pieri}, {Frank}, {Mathur}, {Weinberg}, {York},
  \& {Oppenheimer}}]{Pieri10}
{Pieri}, M.~M., {Frank}, S., {Mathur}, S., {Weinberg}, D.~H., {York}, D.~G., \&
  {Oppenheimer}, B.~D., 2010, \apj, 716, 1084

\bibitem[{{Reimers} {et~al.}(1997){Reimers}, {Kohler}, {Wisotzki}, {Groote},
  {Rodriguez-Pascual}, \& {Wamsteker}}]{Reimers97}
{Reimers}, D., {Kohler}, S., {Wisotzki}, L., {Groote}, D., {Rodriguez-Pascual},
  P., \& {Wamsteker}, W., 1997, \aap, 327, 890

\bibitem[{{Scannapieco} {et~al.}(2006){Scannapieco}, {Pichon}, {Aracil},
  {Petitjean}, {Thacker}, {Pogosyan}, {Bergeron}, \&
  {Couchman}}]{Scannapieco06}
{Scannapieco}, E., {Pichon}, C., {Aracil}, B., {Petitjean}, P., {Thacker},
  R.~J., {Pogosyan}, D., {Bergeron}, J., \& {Couchman}, H.~M.~P., 2006, \mnras,
  365, 615

\bibitem[{{Schaye} {et~al.}(2003){Schaye}, {Aguirre}, {Kim}, {Theuns}, {Rauch},
  \& {Sargent}}]{Schaye03}
{Schaye}, J., {Aguirre}, A., {Kim}, T.-S., {Theuns}, T., {Rauch}, M., \&
  {Sargent}, W.~L.~W., 2003, \apj, 596, 768

\bibitem[{{Shull} {et~al.}(2004){Shull}, {Tumlinson}, {Giroux}, {Kriss}, \&
  {Reimers}}]{Shull04}
{Shull}, J.~M., {Tumlinson}, J., {Giroux}, M.~L., {Kriss}, G.~A., \& {Reimers},
  D., 2004, \apj, 600, 570

\bibitem[{{Shull} {et~al.}(2010){Shull}, {France}, {Danforth}, {Smith}, \&
  {Tumlinson}}]{Shull10}
{Shull}, M., {France}, K., {Danforth}, C., {Smith}, B., \& {Tumlinson}, J.,
  2010, arXiv:1008.2957

\bibitem[{{Simcoe} {et~al.}(2004){Simcoe}, {Sargent}, \& {Rauch}}]{Simcoe04}
{Simcoe}, R.~A., {Sargent}, W.~L.~W., \& {Rauch}, M., 2004, \apj, 606, 92

\bibitem[{{Smette} {et~al.}(2002){Smette}, {Heap}, {Williger}, {Tripp},
  {Jenkins}, \& {Songaila}}]{Smette02}
{Smette}, A., {Heap}, S.~R., {Williger}, G.~M., {Tripp}, T.~M., {Jenkins},
  E.~B., \& {Songaila}, A., 2002, \apj, 564, 542

\bibitem[{{Songaila} \& {Cowie}(1996)}]{Songaila96}
{Songaila}, A. \& {Cowie}, L.~L., 1996, \aj, 112, 335

\bibitem[{{Srianand} \& {Petitjean}(2000)}]{Srianand00apm}
{Srianand}, R. \& {Petitjean}, P., 2000, \aap, 357, 414

\bibitem[{{Stumpff}(1980)}]{Stumpff80}
{Stumpff}, P., 1980, \aaps, 41, 1

\bibitem[{{Tripp} {et~al.}(2008){Tripp}, {Sembach}, {Bowen}, {Savage},
  {Jenkins}, {Lehner}, \& {Richter}}]{Tripp08}
{Tripp}, T.~M., {Sembach}, K.~R., {Bowen}, D.~V., {Savage}, B.~D., {Jenkins},
  E.~B., {Lehner}, N., \& {Richter}, P., 2008, \apjs, 177, 39

\bibitem[{{Worseck} {et~al.}(2007){Worseck}, {Fechner}, {Wisotzki}, \&
  {Dall'Aglio}}]{Worseck07}
{Worseck}, G., {Fechner}, C., {Wisotzki}, L., \& {Dall'Aglio}, A., 2007, \aap,
  473, 805

\bibitem[{{Zheng} {et~al.}(2004){Zheng}, {Kriss}, {Deharveng}, {Dixon}, {Kruk},
  {Shull}, {Giroux}, {Morton}, {Williger}, {Friedman}, \& {Moos}}]{Zheng04}
{Zheng}, W., {Kriss}, G.~A., {Deharveng}, J., {et~al.}, 2004, \apj, 605, 631

\end{thebibliography}
\end{document}